\documentclass[twocolumn]{aastex63}

\accepted{\today}
\submitjournal{ApJ}

\shorttitle{A Young Molecular Outflow Driven by a Decelerating Jet}
\shortauthors{Tafoya et al.}
\usepackage{xcolor}
\usepackage{CJK}

\begin{document}
\begin{CJK*}{UTF8}{min}

\title{The ALMA Survey of 70$\mu$m Dark High-mass Clumps in Early Stages (ASHES)\\ III.  A Young Molecular Outflow Driven by a Decelerating Jet}

\correspondingauthor{Daniel Tafoya}
\email{daniel.tafoya@chalmers.se}

\author[0000-0002-2149-2660]{Daniel Tafoya (多穂谷)}
\affiliation{Department of Space, Earth and Environment, Chalmers University of Technology, \\
	Onsala Space Observatory, 439~92 Onsala, Sweden}

\author{Patricio Sanhueza}
\affiliation{National Astronomical Observatory of Japan, National Institutes of Natural Sciences, 2-21-1 Osawa, Mitaka, Tokyo 181-8588, Japan}
\affiliation{Department of Astronomical Science, SOKENDAI (The Graduate University for Advanced Studies), 2-21-1 Osawa, Mitaka, Tokyo 181-8588, Japan}

\author{Qizhou Zhang}
\affiliation{Center for Astrophysics $|$ Harvard \& Smithsonian, 60 Garden Street, Cambridge, MA 02138, USA}

\author{Shanghuo Li}
\affil{Shanghai Astronomical Observatory, Chinese Academy of Sciences, 80 Nandan Road, Shanghai 200030, China}
\affiliation{Center for Astrophysics $|$ Harvard \& Smithsonian, 60 Garden Street, Cambridge, MA 02138, USA}
\affiliation{University of Chinese Academy of Sciences, 19A Yuquanlu, Beijing 100049, China}

\author{Andr\'es E. Guzm\'an}
\affil{National Astronomical Observatory of Japan, National Institutes of Natural Sciences, 2-21-1 Osawa, Mitaka, Tokyo 181-8588, Japan}

\author{Andrea Silva}
\affil{National Astronomical Observatory of Japan, National Institutes of Natural Sciences, 2-21-1 Osawa, Mitaka, Tokyo 181-8588, Japan}

\author{Eduardo de la Fuente}
\affil{Departamento de F\'{i}sica, CUCEI, Universidad de Guadalajara, Blvd. Gral. Marcelino Garc\'{i}a Barrag\'an 1421, Ol\'impica, 44430, Guadalajara, Jalisco, M\'exico}.
\affil{Information Technologies Ph. D., CUCEA, Universidad de Guadalajara Perif\'erico Norte 799, N\'ucleo Universitario Los Belenes, 45100 Zapopan, Jalisco, M\'exico.}

\author{Xing Lu}
\affil{National Astronomical Observatory of Japan, National Institutes of Natural Sciences, 2-21-1 Osawa, Mitaka, Tokyo 181-8588, Japan}

\author{Kaho Morii}
\affil{Department of Astronomy, Graduate School of Science, The University of Tokyo, 7-3-1 Hongo, Bunkyo-ku, Tokyo 113-0033, Japan}
\affiliation{National Astronomical Observatory of Japan, National Institutes of Natural Sciences, 2-21-1 Osawa, Mitaka, Tokyo 181-8588, Japan}

\author{Ken'ichi Tatematsu}
\affil{National Astronomical Observatory of Japan, National Institutes of Natural Sciences, 2-21-1 Osawa, Mitaka, Tokyo 181-8588, Japan}

\author{Yanett Contreras}
\affil{Leiden Observatory, Leiden University, PO Box 9513, NL-2300 RA Leiden, the Netherlands}

\author{Natsuko Izumi}
\affil{National Astronomical Observatory of Japan, National Institutes of Natural Sciences, 2-21-1 Osawa, Mitaka, Tokyo 181-8588, Japan}
\affil{College of Science, Ibaraki University, 2-1-1 Bunkyo, Mito, Ibaraki 310-8512, Japan}

\author{James M. Jackson}
\affil{SOFIA Science Center, USRA, NASA Ames Research Center, Moffett Field CA 94045, USA}

\author{Fumitaka Nakamura}
\affil{National Astronomical Observatory of Japan, National Institutes of Natural Sciences, 2-21-1 Osawa, Mitaka, Tokyo 181-8588, Japan}
\affiliation{Department of Astronomical Science, SOKENDAI (The Graduate University for Advanced Studies), 2-21-1 Osawa, Mitaka, Tokyo 181-8588, Japan}

\author{Takeshi Sakai}
\affil{Graduate School of Informatics and Engineering, The University of Electro-Communications, Chofu, Tokyo 182-8585, Japan.}

%\affiliation{}



\begin{abstract}

We present a spatio-kinematical analysis of the CO~($J$=2$\rightarrow$1) line emission, observed with the Atacama Large Millimter/submillimter 
Array (ALMA), of the outflow associated with the most massive core, ALMA1, in the 70 $\mu$m dark clump G010.991$-$00.082. The 
position-velocity (P-V) diagram of the molecular outflow exhibits a peculiar $\mathsf{S}$-shaped morphology that has not been seen in any other star 
forming region. We propose a spatio-kinematical model for the bipolar molecular outflow that consists of a decelerating high-velocity component surrounded
by a slower component whose velocity increases with distance from the central source. The physical interpretation of the model is in terms of a jet that 
decelerates as it entrains material from the ambient medium, which has been predicted by calculations and numerical simulations of molecular outflows in the 
past. One side of the outflow is shorter and shows a stronger deceleration, suggesting that the medium through which 
the jet moves is significantly inhomogeneous. The age of the outflow is estimated to be $\tau$$\approx$1300 years, after correction for a mean 
inclination of the system of $\approx$57$^{\circ}$.

\end{abstract}

\keywords{stars: protostars --- stars: jets --- ISM: jets and outflows --- techniques: imaging spectroscopy}


\section{Introduction} \label{sec:intro}

The jet phenomenon is ubiquitous in the universe as it is found in many astrophysical contexts over a wide range of spatial scales. 
Particularly, in star-forming regions jets can drive massive molecular outflows that reveal the presence of young protostellar objects that 
are still accreting material from their parent clouds \citep[see e.g.,][]{Arce2007,Bally2016,Lee2020}. Molecular outflows in high-mass 
star-forming regions have been found from the earliest stages of evolution in infrared dark clouds \citep[IRDCs;][]{Sanhueza2010, Wang2011,
Sakai2013, Lu2015, Zhang2015,Kong2019, Li2019,Svoboda2019} to later stages with evident signs of star formation 
\citep[e.g.,][]{Beuther2002,Zhang2005,Qiu2008,Yang2018,Li2019a,Nony2020,Li2020}. 
The study of jets and molecular outflows in the context of star formation is important since they provide crucial information on the accretion 
history of the central object. In addition, jets are thought to facilitate the accretion process by removing angular momentum from the disk 
and, eventually, they may also play a role in quenching the accretion by dispersing the material of the parent cloud. Furthermore, the study 
of the physical processes behind the launching and collimation of jets in star formation is important in itself since it can contribute 
to better understand the jet phenomenon in other astrophysical contexts.

A powerful tool that is commonly used to study the spatio-kinematical characteristics of jets and their associated molecular outflows is 
the position-velocity (P-V) diagram. The P-V diagrams of the outflows of low- and high-mass star-forming regions have revealed the 
presence of different components with specific kinematical signatures. Particularly, it has been found that many outflows exhibit components whose 
velocity increases with distance (so-called ``Hubble-law''). The Hubble-law may appear associated with several components in such a way that they form a 
jagged profile, which is referred to as Hubble wedge. The tips of the spurs of such Hubble wedge are sometimes identified as discrete components, called knots, 
whose velocity decreases with distance from the central source 
\citep[e.g., L1448, HH~211, CARMA-7, W43-MM1($\#67$);][]{Bachiller1990,Palau2006,Hirano2006,Plunkett2015, Nony2020}. These particular velocity profiles are 
explained in terms of internal shocks, within a collimated outflow or jet, produced by variations in the mass-loss rate, which in turn are thought to be due to variations 
in the mass-accretion rate of the central source. \cite{Arce2007} summarized the morphologies of the spatial distributions and P-V diagrams obtained 
from models of different types of outflows. In general, the morphology of the outflows and the shape of their corresponding P-V diagrams depend on the specific 
details of the geometry and physical conditions of the jet as well as of those of the ambient medium. Thus, it is important to carry out observations of molecular 
outflows to characterize their physical conditions and constrain the values of the input parameters of the models.   

\begin{figure*}[!t]
	\centering
	\includegraphics[angle=0,scale=1.1]{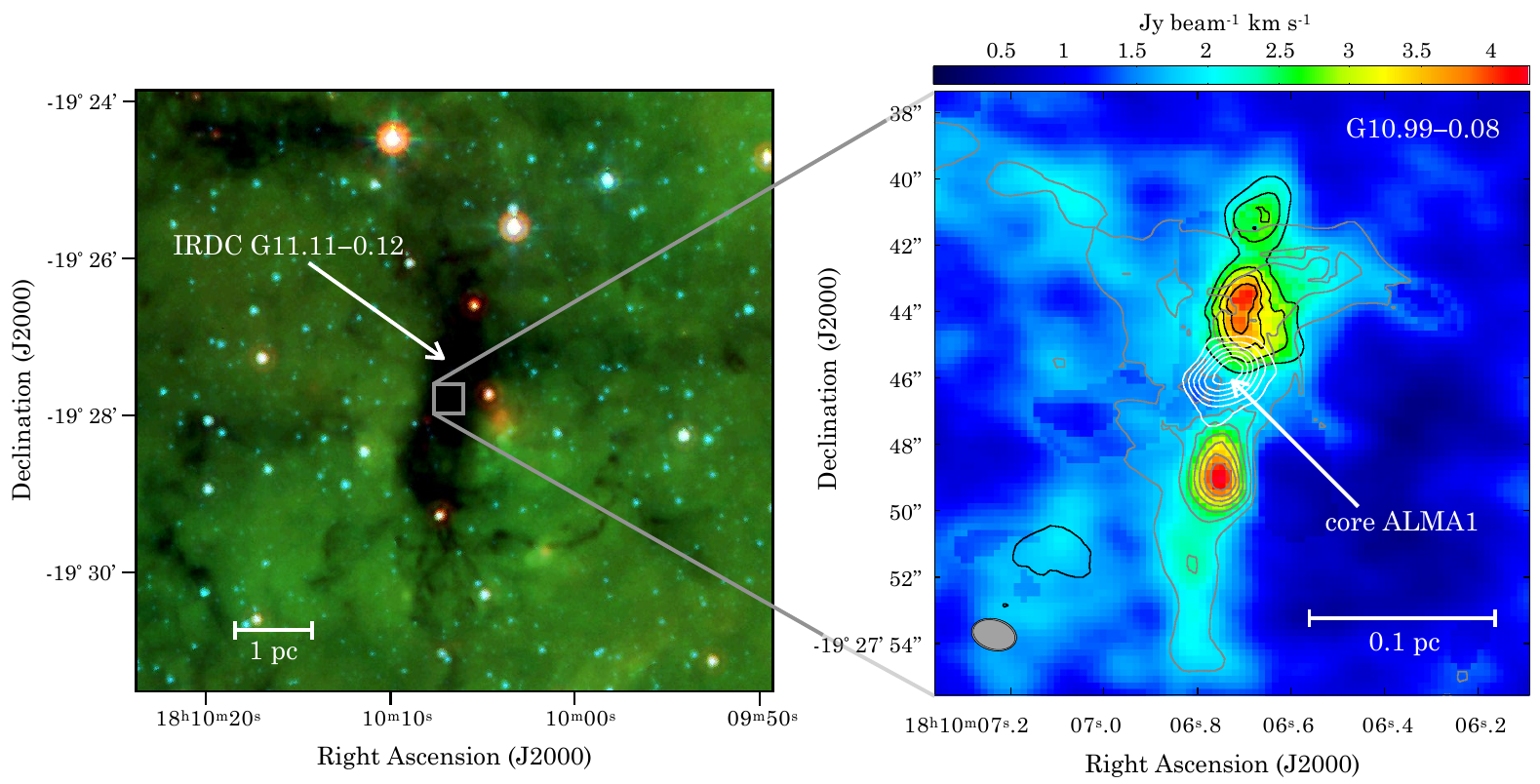}
	\caption{Locations of the IRDC G11.11$-$0.12, the clump G10.99$-$0.08 and the core ALMA1. {\bf Left panel:} Three-colour view of the vicinity of the 
		IRDC G11.11$-$0.12  obtained with Spitzer IRAC and MIPS data with 5.8 $\mu$m (blue), 8.0 $\mu$m (green), 
		24.0 $\mu$m (red) from the GLIMPSE and MIPSGAL survey \citep{Churchwell2009,Carey2009}. {\bf Right panel:} Moment-0 of the 
		CO($J$=2$\rightarrow$1) emission integrated over the velocity range  $-28$$<$v$_{\rm LSRK}$(km~s$^{-1}$)$<$+104.  The image 
		includes only pixels with a brightness in the range 4-250~mJy~beam$^{-1}$. The white contours indicate the 1.3~mm continuum emission. The contours are 
		rms$\times$2$\times$$ i$, with rms=$1.5\times10^{-4}$~Jy~beam$^{-1}$ and $i$=3,4,5$\ldots$  The black and grey contours represent velocity-integrated 
		emission in the velocity range  $-28$$<$v$_{\rm LSRK}$(km~s$^{-1}$)$<$+26 and +26$<$v$_{\rm LSRK}$(km~s$^{-1}$)$<$+104, respectively. 
		The contours are rms$\times$2$\times$$ i$, with rms=$1.8\times10^{-1}$~Jy~beam$^{-1}$~km~s$^{-1}$ and $i$=3,4,5$\ldots$
		The filled ellipse located at the bottom--left corner represents the size of the synthesized beam, 
		$\theta_{\rm FWHM}$=1$\rlap{.}^{\prime\prime}$38$\times$0$\rlap{.}^{\prime\prime}$95, P.A.=76$^{\circ}$. }
	\label{Fig1a}%
\end{figure*}

In the ALMA Survey of 70 $\mu$m Dark High-mass Clumps in Early Stages \citep[ASHES:][]{Sanhueza2019} we investigate the early stages 
of high-mass star formation using the Atacama Large Millimeter/submillimeter Array (ALMA). In a pilot survey, we carried out high-angular resolution 
observations towards 12 massive 70 $\mu$m dark clumps. The sample was selected by combining the ATLASGAL survey
\citep{Schuller2009,Contreras2013} and a series of studies from the MALT90 survey \citep{Foster2011,Sanhueza2012,Foster2013,Jackson2013,Guzman2015,
Rathborne2016,Contreras2017,Whitaker2017}. The source selection and fragmentation properties of the sample are described in detail by \cite{Sanhueza2019}. 

One of the ASHES targets is the clump G010.991$-$00.082 (hereafter, G10.99$-$0.08), which is located at the distance of 3.7 kpc 
\citep[][and references therein]{Pillai2006,Henning2010,Kainulainen2013,Wang2016,Sanhueza2019,Pillai2019}. This clump has no point sources 
detected in the near or mid-infrared either in the GLIMPSE nor in the MIPS Galactic Plane Survey \citep{Churchwell2009,Carey2009}. Neither does it have 
point sources detected at 70 $\mu$m in the HiGAL survey \citep{Molinari2010}, indicating that this clump is at a very early stage of its evolution 
\citep[e.g.][]{Sanhueza2013,Tan2013,Sanhueza2017}. A SED fitting for G10.99$-$0.08 gives a dust temperature of 12~K and a mass of 1810 
$M_{\odot}$ \citep{Sanhueza2019}. Recently, \cite{Pillai2019}  presented SMA observations toward G10.99$-$0.08 and identified 
structures that seem to be molecular outflows driven by low-mass protostars. \cite{Li2020} confirmed the presence of blue- and red-shifted emission associated 
with the brightest core (ALMA1; $M\sim10 M_{\odot}$) suggesting that it is indeed driving a bipolar outflow, although, given the complexity of the emission in the 
region, it was not possible to characterise in detail the observed structures. Nevertheless, since G10.99$-$0.08 is a relatively young star-forming region, it is 
an attractive target that deserves further investigation to study the first stages of development of molecular outflows in a relatively massive core that has the 
potential to form a high-mass star in the future. 
 
As part of the series of papers derived from the ASHES survey, here we present an analysis of  the CO~($J$=2$\rightarrow$1) line emission observed with 
ALMA toward G10.99$-$0.08 to study in detail the morphology and kinematics of the bipolar molecular outflow associated with core ALMA1, whose P-V diagram 
exhibits a peculiar $\mathsf{S}$-shaped morphology that has not 
been seen in other molecular outflows before. A summary of the details of the observations used in this work is presented in \S\ref{observations}.  The description 
of the analysis of the data and the presentation of a spatio-kinematical model that explains the observations are given in section \S\ref{results}. 
In section \S\ref{discussion} we discuss the physical interpretation of the spatio-kinematical model and estimate the age and energetics of the molecular 
	outflow. In this section we also discuss on the nature of the driving jet. Finally, the conclusions of this work are presented in 
section \S\ref{conclusions}

\begin{figure*}[!t]
	\centering
	\includegraphics[angle=0,scale=1.0]{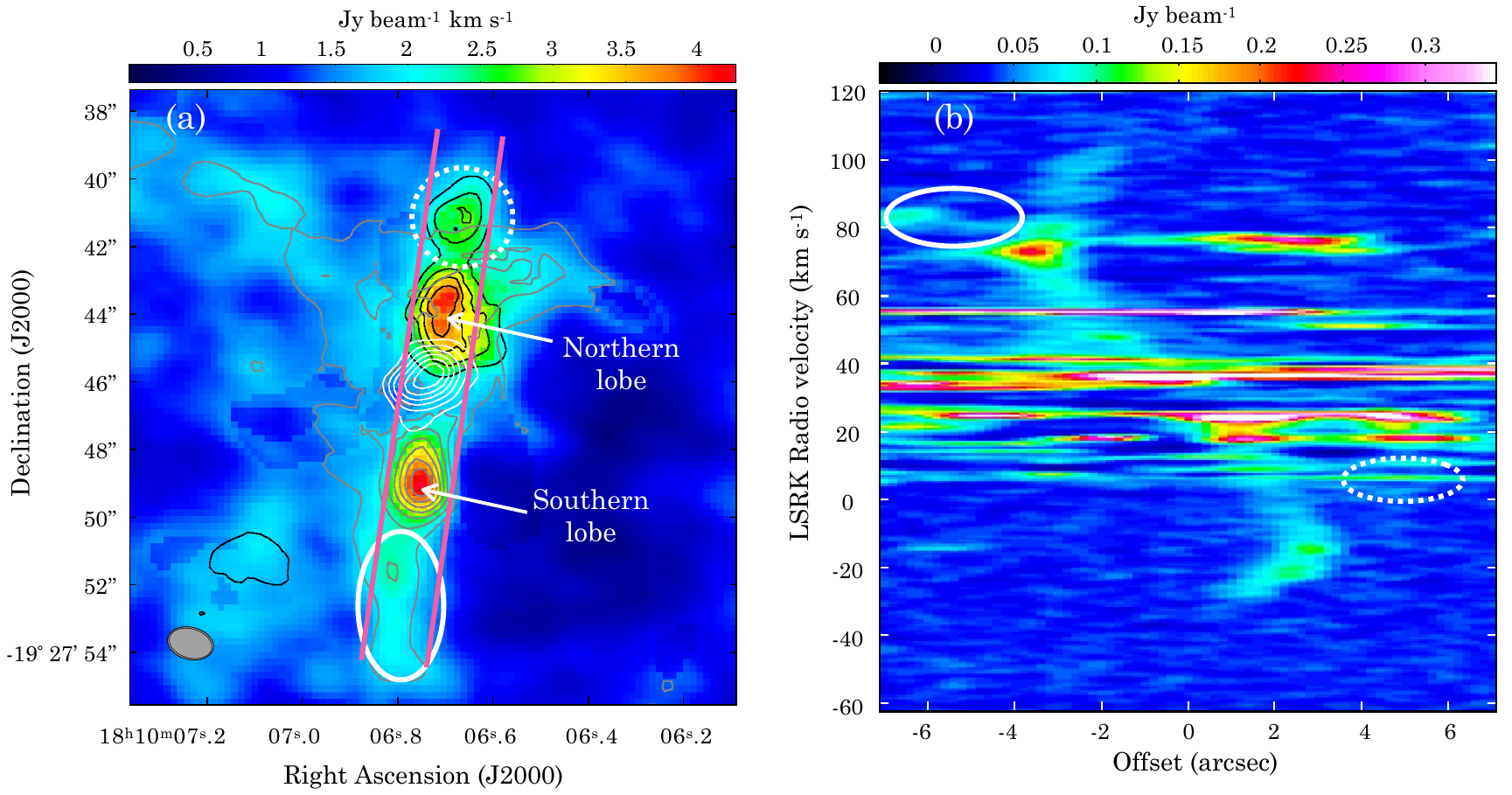}
	\caption{ALMA observations of the CO($J$=2$\rightarrow$1) emission in G10.99$-$0.08. {\bf (a)} Same as right panel of Fig.~\ref{Fig1a}. 
		{\bf (b)} P-V diagram of the CO($J$=2$\rightarrow$1) emission obtained by using the slit indicated with parallel 
		pink lines shown in the left panel. The horizontal axis is the position offset with respect to the position of the continuum peak of the 
		core ALMA1, (J2000) R.A.=18$^{\rm h}$10$^{\rm m}$06$\rlap{.}^{\rm s}$736, Dec.=$-$19$^{\circ}$27$^{\prime}$45$\rlap{.}^{\prime\prime}$88. 
		The ellipses indicate emission aligned in the direction connecting the northern and southern lobes but that does not follow the $\mathsf{S}$-shaped 
		pattern associated with the bright lobes (see main text).}
	\label{Fig1b}%
\end{figure*}

\section{Observations}\label{observations}

The observations used in the present analysis were carried out on January 28, 2016 using 41 antennas of the ALMA 12m array with Band 6 receivers 
($\sim$224 GHz; 1.34 mm) as part of the project 2015.1.01539.S (P.I.: P. Sanhueza). The array was arranged in configuration C36-1 and the maximum 
and minimum baseline lengths were 330~m and 15~m, respectively. The corresponding angular resolution and maximum recoverable scale are 
0$\rlap{.}^{\prime\prime}$95 and 8$\rlap{.}^{\prime\prime}$8, respectively. The data were calibrated manually with CASA using J1924-2914 (3.05 Jy) 
and J1733-1304 (1.65~Jy) as bandpass and gain calibrators, respectively. The flux was calibrated using Titan. The continuum emission was produced 
by averaging line-free channels in visibility space \cite[see additional details for the continuum in][]{Sanhueza2019}. Channel maps of the 
CO~($J$=2$\rightarrow$1) line with a spectral resolution of $\sim$1.3 km~s$^{-1}$ were created after subtracting the continuum emission 
from the data cubes. The imaging of the channel maps was performed setting the multi-scale option value to 0, 5, 15, and 25 times the 
size of the pixel of 0$\rlap{.}^{\prime\prime}$2. The selection of masks for cleaning was done automatically using the cleaning algorithm {\sc yclean} developed 
by \cite{Contreras2018a}. The Briggs weighting robust parameter was set to 0.5, which resulted in a final angular resolution for the images 
of 1$\rlap{.}^{\prime\prime}$38$\times$0$\rlap{.}^{\prime\prime}$95, PA=76$^{\circ}$ ($\sim$0.025~pc at 3.7~kpc). The typical channel 
root-mean-square (rms) noise level in the resulting channel maps is $\sim$4~mJy~beam$^{-1}$.

\section{Data analysis  and Results}\label{results}

The clump G10.99$-$0.08 is embedded in the IRDC G11.11$-$0.12, as it is shown in Fig.~\ref{Fig1a}. \cite{Pillai2019} presented observations carried out with 
the SMA toward G10.99$-$0.08, with an angular resolution of $\sim$4$^{\prime\prime}$ ($\sim$0.07~pc at 3.7~kpc), and found hints of the presence of 
a molecular outflow in this clump. Particularly, in their Figure 2 there is blue- and red-shifted CO($J$=2$\rightarrow$1) line emission distributed in a more or 
less bipolar fashion and centred at the position of the brightest 1.3~mm continuum peak (core ALMA1). \cite{Li2020} 
confirmed the presence of such a bipolar structure, which they identified as a bipolar outflow associated  with core ALMA1. 
The ALMA observations of the CO($J$=2$\rightarrow$1) line emission presented in this work clearly reveal two bright lobes located toward the north and south 
of core ALMA1 (see right panel of Fig.~\ref{Fig1a}). 

The colour map in the right panel of Fig.~\ref{Fig1a} is a moment-0 image of the emission integrated over the velocity range $-28$$<$v$_{\rm LSRK}$(km~s$^{-1}$)$<$+104, which, as it is shown below, includes all the CO($J$=2$\rightarrow$1) emission associated to the bipolar outflow, 
and the white contours indicate the 1.3~mm continuum emission. The moment-0 image was created using only pixels with a brightness in the range 4-250~mJy~beam$^{-1}$  
to enhance the emission of the bipolar outflow, nonetheless some extended emission from the ambient medium is also visible in the image. The black and grey 
contours represent velocity-integrated emission in the velocity range $-28$$<$v$_{\rm LSRK}$(km~s$^{-1}$)$<$+26 and +26$<$v$_{\rm LSRK}$(km~s$^{-1}$)$<$+104, 
respectively. The derived systemic velocity of the bipolar outflow is v$_{\rm sys, LSRK}$=26.1~km~s$^{-1}$ (see below). Thus, the gas of the northern lobe 
is moving toward us, while the southern lobe is moving away from us. 

\begin{figure*}
	\centering
	\includegraphics[angle=-0,scale=0.45]{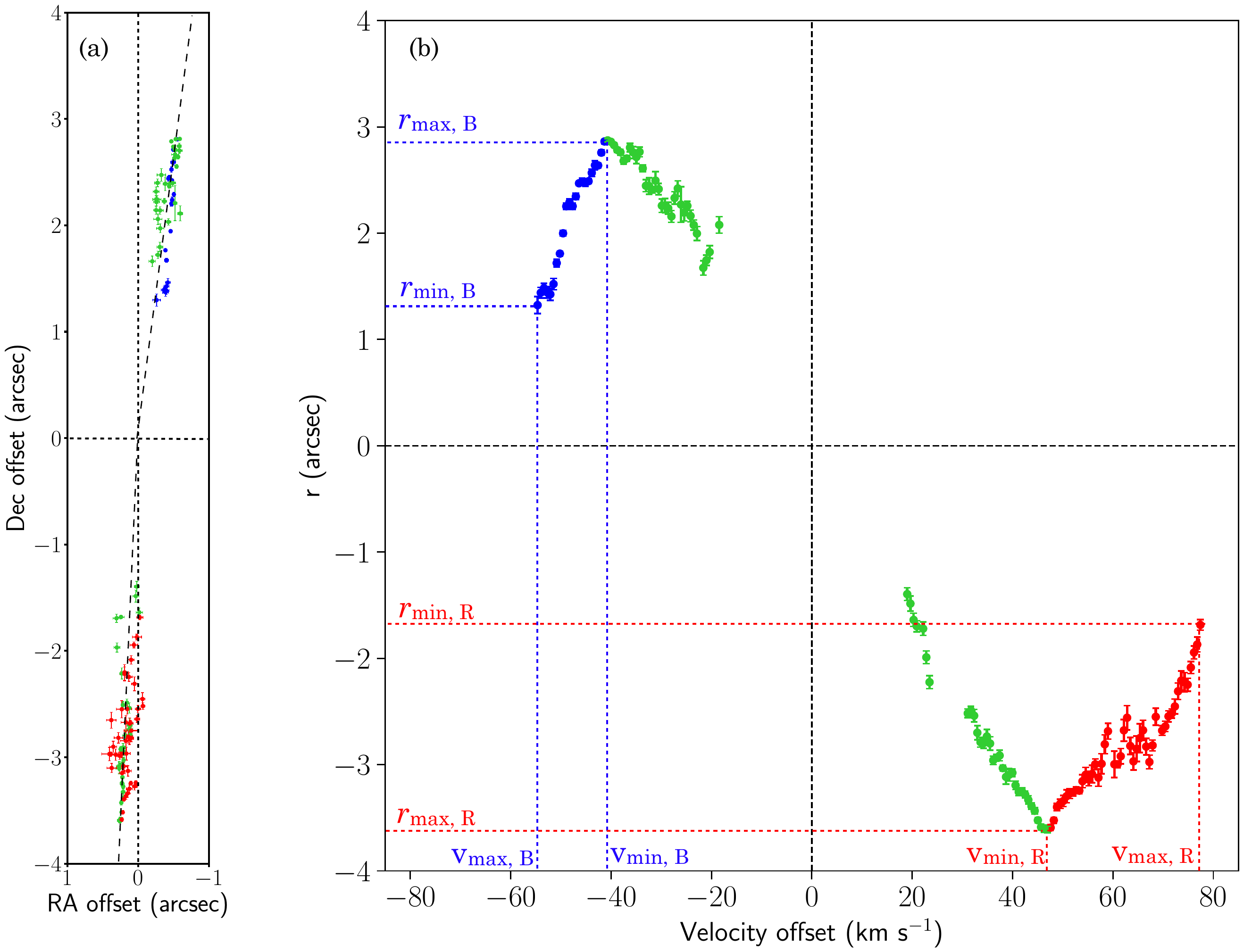}
	\caption{Spatial and velocity distribution of the CO($J$=2$\rightarrow$1) emission in G10.99$-$0.08. {\bf (a)} Peak positions of the  
		CO($J$=2$\rightarrow$1) emission for 
		each individual velocity channel obtained by fitting a 2D-Gaussian function. The origin corresponds to the peak position 
		of the continuum emission (see main text). The negative R.A. offsets are on the 
		right side of the plot. {\bf (b)} Position offsets from peak position of the continuum as a function of the velocity offset 
		from the systemic velocity v$_{\rm sys, LSRK}$=26~km~s$^{-1}$. For both panels, the colour of the data points 
		are coded according to their velocity gradient, d$|$v$|$/d$|r|$; green for d$|$v$|$/d$|r|$$>$0 and blue-red
		for d$|$v$|$/d$|r|$$<$0. The negative distance offsets correspond to data points with negative Declination offsets. The 
		error bars represent the nominal errors from the 2D-Gaussian fit.}
	\label{Fig2}%
\end{figure*}

In order to scrutinize the spatial distribution and kinematics of the bipolar lobes, we created a P-V diagram using the slit indicated with parallel 
pink lines in Fig.~\ref{Fig1b}a. The resulting P-V diagram is shown in Fig.~\ref{Fig1b}b. The vertical axis is the LSRK radio velocity 
and the horizontal axis is the position offset with respect to the position of the 1.3~mm continuum peak, 
(J2000) R.A.=18$^{\rm h}$10$^{\rm m}$06$\rlap{.}^{\rm s}$736, Dec.=$-$19$^{\circ}$27$^{\prime}$45$\rlap{.}^{\prime\prime}$88. The 
P-V diagram reveals an $\mathsf{S}$-shaped feature crossed by several horizontal components. Inspection of individual channels in the data cube 
shows that the horizontal components are due to extended emission of the ambient gas. On the other hand, the $\mathsf{S}$-shaped feature is 
associated with emission from the bipolar lobes seen in Fig.~\ref{Fig1b}a. It is worth noting that there is emission that extends northward and southward of 
the two bright lobes, which is indicated in Fig.~\ref{Fig1b}a with a dashed and a solid oval, respectively. The corresponding emission in 
the P-V diagram is also indicated with a dashed and a solid oval (see Fig.~\ref{Fig1b}b). It can be seen that such emission does not follow 
the $\mathsf{S}$-shaped pattern associated with the bright lobes, implying that it is most likely arising in the gas of the 
ambient medium.   

Interestingly, the $\mathsf{S}$-shaped pattern in the P-V diagram of Fig.~\ref{Fig1b}b resembles the P-V diagram of the evolved star 
IRAS 16342$-$3814 \citep{Sahai2017, Tafoya2019}.  Given the similarity of those two P-V diagrams, we undertook an analysis of data of 
G10.99$-$0.08 adopting the approach taken by \cite{Tafoya2019}, which is described in the following.

Firstly, from the data cube of the CO($J$=2$\rightarrow$1) line we measured the flux density and peak position 
of the emission associated with the bipolar lobes in every single channel within the velocity range $-28$$<$v$_{\rm LSRK}$(km~s$^{-1}$)$<$+104, 
which includes all the emission of the $\mathsf{S}$-shaped pattern of the P-V diagram, except the channels with significant contamination 
from extended emission. Given that the emission in individual channels is not well resolved, 
we fitted a 2D-Gaussian function to obtain the peak positions. Subsequently, we calculated the separation of the emission 
peak positions from the peak position of the 1.3~mm continuum, which is defined as the reference 
position. In Fig.~\ref{Fig2}a we plot the declination (Dec.) and right ascension (R.A.) offsets of the emission peak positions with 
respect to the reference position. The points appear clustered in two elongated structures that
correspond to the northern and southern lobes. A linear fit to the distribution of points in each of the lobes reveals that while the points 
of the northern lobe lie on a line with P.A.=$-$8$^{\circ}$, the points of the southern lobe lie on a line with P.A.=$-$2$^{\circ}$. 
In Fig.~\ref{Fig2}b we plot the distance of the emission peak from the reference position, $r=\pm({\rm Dec.offset}^{2}+{\rm R.A.offset}^{2})^{1/2}$, 
as a function of the velocity offset with respect to the reference velocity. We took the negative solution of the square root for data points whose 
Dec. offset is negative. Initially, the systemic velocity reported in previous works \citep{Sanhueza2019,Pillai2019} was used as the reference velocity 
to calculate the velocity offsets. However, the reference velocity was later redefined in such a way that a linear fit of the green points in 
Fig.~\ref{Fig2}b (see below for the definition of the colouring code) passes through the origin of the plot. The resulting reference velocity has a 
value of v$_{\rm ref, LSRK}$=26.1~km~s$^{-1}$, and it is defined as the systemic velocity of the outflow. The physical justification for 
interpreting the reference velocity in this manner is further discussed in \S\ref{discussion}. 

\begin{figure*}
	\centering
	\includegraphics[angle=0,scale=0.69]{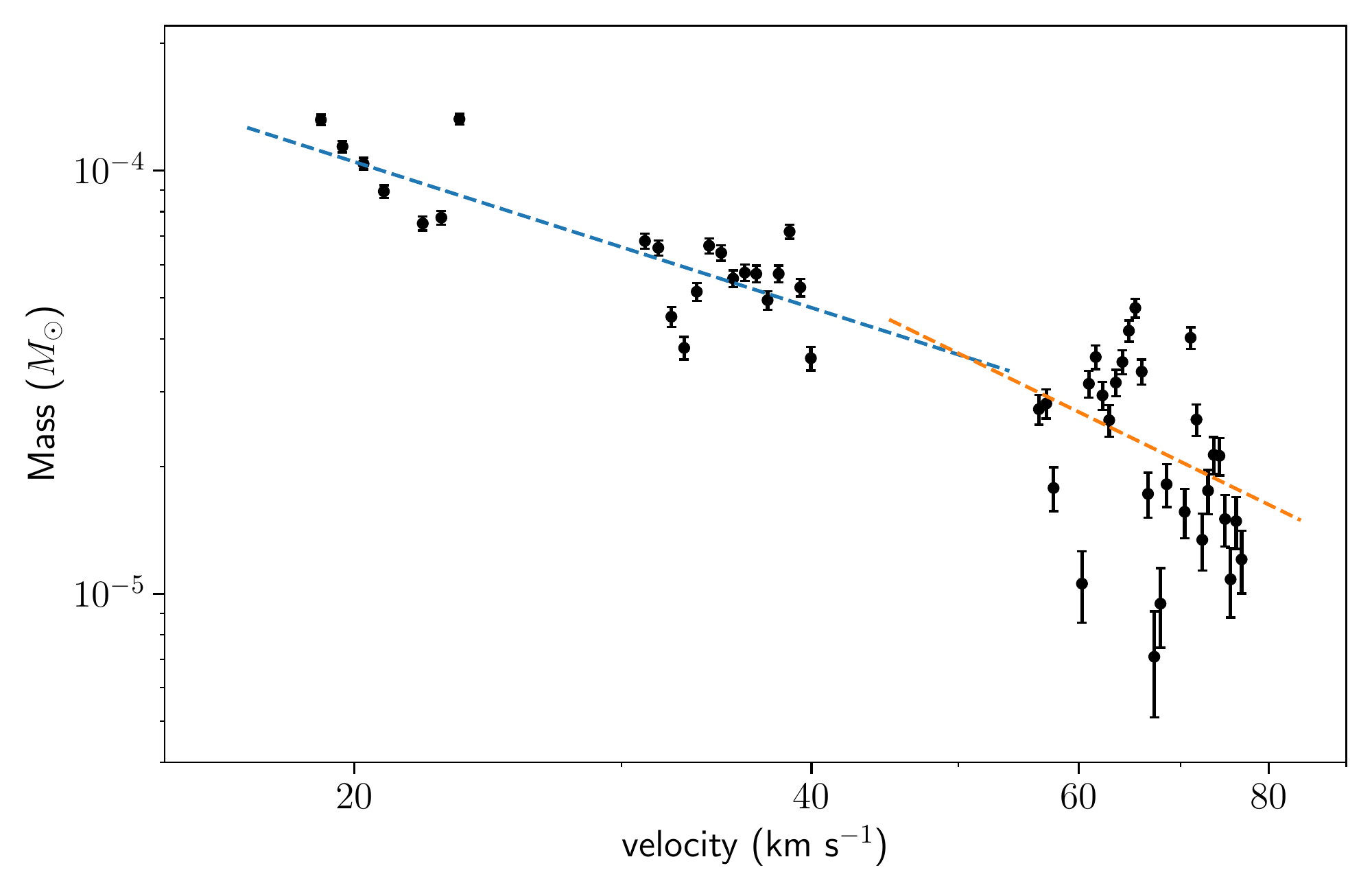}
	\caption{Molecular mass as a function of expansion velocity (mass spectrum) obtained from the emission that is red-shifted with respect to the 
		systemic velocity. The error bars were obtained using the rms noise of the individual channels of 4~mJy beam$^{-1}$.
		The light-blue and orange dashed lines are power-law fits to the data in the velocity ranges 10$<$v$_{\rm offset}$(km~s$^{-1}$)$<$50, 
			and 50$<$v$_{\rm offset}$(km~s$^{-1}$)$<$80, respectively.}
	\label{Fig3}%
\end{figure*}

In a similar way to the results obtained by \cite{Tafoya2019}, the data points shown in Fig.~\ref{Fig2}b  reveal two kinematical 
components: one with intermediate velocities offsets, in the range $-41$$<$v$_{\rm offset}$(km~s$^{-1}$)$<$+48, and whose 
velocity gradient is positive, i.e., d$|$v$|$/d$|$$r$$|$ $>$ 0; and another component with higher velocity offsets, in the 
ranges $-54$$<$v$_{\rm offset}$(km~s$^{-1}$)$<$$-41$ and +48$<$v$_{\rm offset}$(km~s$^{-1}$)$<$+78, and whose 
velocity gradient is negative. The data points in Fig.~\ref{Fig2}b are colour-coded according to their velocity 
gradient: green for d$|$v$|$/d$|$$r$$|$$>$0 and blue or red for d$|$v$|$/d$|$$r$$|$$<$0\footnote{The colouring code adopted here is 
the same as the one used by \cite{Tafoya2019}. The blue and red colours of the component with d$|$v$|$/d$|$$r$$|$$<$0 indicate 
that the emission is blue- and red-shifted with respect to the reference velocity. However, for the component with d$|$v$|$/d$|$$r$$|$$>$0 
only green colour is used.}. In accordance with \cite{Tafoya2019}, the kinematical component with positive velocity gradient is referred to as 
``low velocity component'' (LVC), since it has relatively lower velocities offsets, and the component with negative velocity gradient, which has higher velocity offsets, 
is referred to as ``high velocity component'' (HVC). It is worth noting that, although the LVC is represented with green colour, it contains material that is 
both blue- and red-shifted with respect to the systemic velocity. From Fig.~\ref{Fig2}a it can be seen that, similarly to IRAS~16342$-$3814 
\cite[see Figure 1 of][]{Tafoya2019}, despite the different kinematical signatures of the LVC and HVC, they appear aligned in the same direction on the sky. 
On the other hand, the distribution of the data points in Fig.~\ref{Fig2}b exhibits a morphology that resembles the $\mathsf{S}$-shaped pattern of the 
emission in the P-V diagram of Fig.~\ref{Fig1b}b, although rotated 90$^{\circ}$. This is because the  emission peak positions lie basically along one single 
direction, which makes the plot of Fig.~\ref{Fig2}b equivalent to a P-V diagram.

\section{A molecular outflow driven by a decelerating jet from core ALMA1}\label{discussion}

The simplest spatio-kinematical model that, in principle, can explain the morphology of the $\mathsf{S}$-shaped seen in the P-V diagram of Fig.~\ref{Fig1b}b 
considers a single outflow of material expanding with constant velocity and a large precession angle. In this model, the $\mathsf{S}$-shaped pattern 
results from the projection effect of the material expanding along different directions. The resulting morphology of the outflow 
is also an $\mathsf{S}$-shaped bipolar structure with some material moving in directions near the line-of-sight and some other material moving nearly on the plane 
of the sky \citep[e.g.,][]{Sahai2017}. However, Fig.~\ref{Fig2}a shows that the HVC and LVC are aligned along narrow, linear structures. Thus, in order to simultaneously explain 
the morphologies of the P-V diagram and the spatial distribution of the emission, the constant-velocity model would require a precession angle $\sim$90$^{\circ}$, i.e., 
the outflow would need to be precessing on a plane that is oriented almost perfectly perpendicular to the plane of the sky, which is very unlikely. 
Furthermore, this model neglects the fact that the velocity of the gas in the outflow would depend on the hydrodynamic interaction with ambient gas. 

Other typical outflow models, such as the unified wind-driven model that includes a collimated as well as wide opening angle wind \citep[e.g.,][]{Shang2006,Banerjee2006,Machida2008} 
predict P-V diagrams that do not resemble the $\mathsf{S}$-shaped pattern seen in Fig.~\ref{Fig1b}b \citep[see e.g.,][]{Hirano2010}. Similarly, spatio-kinematical 
models such as biconical outflows \citep[e.g.,][]{Cabrit1986,Cabrit1990} and expanding bipolar bubbles \citep[e.g.,][]{Shu1991,Masson1992}, can be easily ruled out 
since the resulting morphologies of their P-V diagram differs from the ones seen in Fig.~\ref{Fig1b}b. 

\cite{Tafoya2019} demonstrated that it is possible to explain the spatial distribution and $\mathsf{S}$-shaped P-V diagram of the CO($J$=2$\rightarrow$1) 
line emission of the evolved star IRAS~16342$-$3814 with a spatio-kinematical model that includes two components: i) a collimated high-velocity component 
that decelerates with distance from the central source and ii) a coaxial slower and less collimated component whose velocity increases with distance from the central source. 
As mentioned in the previous section, \S\ref{results}, the outflow of core ALMA1 has components with the same kinematical characteristics of those of the evolved 
star IRAS~16342$-$3814. In addition, the morphologies of the spatial distributions and the P-V diagrams of both sources exhibit remarkable similarities. 
Thereby, we propose that the CO($J$=2$\rightarrow$1) line emission of the outflow of core ALMA1 can be explained by the spatio-kinematical model presented by 
\cite{Tafoya2019} \footnote{It should be pointed out that the spatio-kinematical model proposed by \cite{Tafoya2019} is not incompatible with the unified wind-driven 
	models \citep{Shang2006}, provided that the wide opening angle wind is not present, or it is too weak to be detected.}.      

\begin{figure*}
	\centering
	\includegraphics[angle=0,scale=0.9]{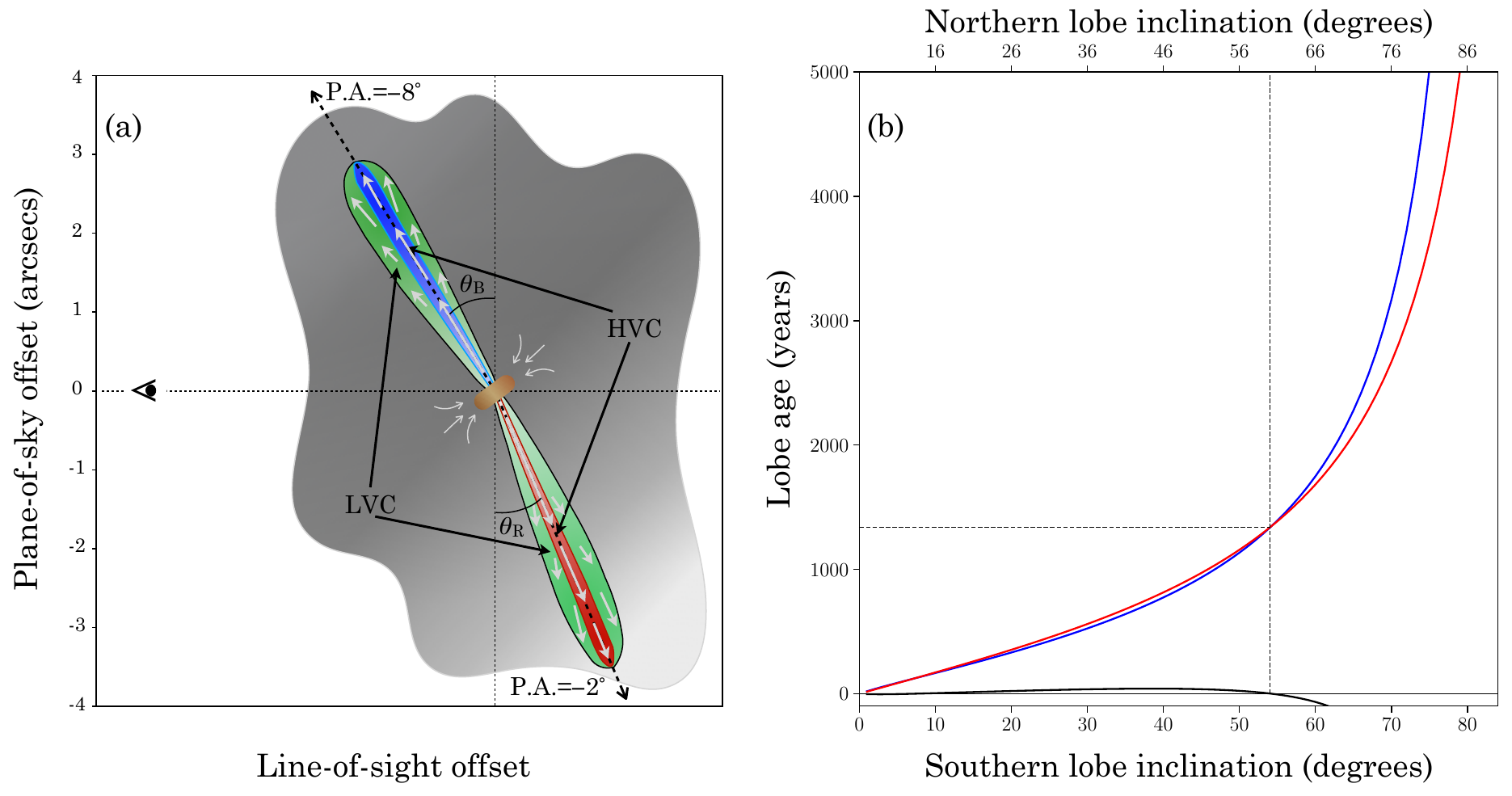}
	\caption{{\bf (a)} Schematic representation of the spatio-kinematical model of the molecular outflow of core ALMA1 in G10.99$-$0.08. 
		The green region indicates the LVC, whose velocity increases with distance from the central source. The blue and red regions indicate 
		the HVC, which decelerates as a function of distance from the central source. The observer is located on the left side of the diagram. {\bf (b)} 
		Kinematical age of the blue- and red-shifted parts of the HVC, as function of their inclination angle, obtained from Equation~\ref{Eq:1}. The 
		blue and red solid lines correspond to the blue- and red-shifted part of the HVC, respectively. The black line indicates the difference between the 
		blue and red lines. The ages were calculated assuming a 
		relative inclination of the lobes of ($\theta_{\rm B}-\theta_{\rm R}$)=6$^{\circ}$. The vertical dashed line indicates the absolute inclination angles 
		of the lobes for which their ages are equal, $\tau$=1300~years. The age is indicated by the horizontal dashed line.}
	\label{Fig4}%
\end{figure*}

The physical interpretation of the spatio-kinematical model that \cite{Tafoya2019} proposed for IRAS~16342$-$3814 is given in terms of a jet-driven molecular 
outflow in which the HVC corresponds to molecular gas entrained along the sides of the jet \citep[steady-state entrainment;][]{De-Young1986,Chernin1994}, 
and the LVC is associated with entrained material that carries momentum transferred through the leading bow shock (prompt entrainment). This interpretation 
is supported by the results of \cite{Smith1997}, who carried out numerical simulations of jet-driven bipolar outflows and obtained two kinematical 
components that correspond to the HVC and LVC seen in IRAS~16342$-$3814 and G10.99$-$0.08 (see their Figure 4). Although it should be noted that
in their simulations the speed of the HVC remains constant across the outflow, thus the P-V diagram has a \reflectbox{$\mathsf{Z}$}-shaped, instead of 
an $\mathsf{S}$-shaped, morphology. Another piece of evidence that is consistent with the jet-driven outflow interpretation is the profile of the 
mass spectrum, $m$(v), since it has been found that such outflows exhibit a power-law variation of mass with velocity, i.e., $m(\rm v)$$\propto$v$^{-\gamma}$ 
\citep[][and references therein]{Raga1993, Lada1996, Cabrit1997}. We obtained the mass spectrum of the outflow of core ALMA1 and it is shown 
in Fig.~\ref{Fig3}. The molecular mass was calculated using the measured flux density of the CO($J$=2$\rightarrow$1) emission from the outflow in channels with 
no contamination by emission of the ambient material. Since the blue-shifted lobe seems to be suffering more contamination by emission of the ambient material, Fig.~\ref{Fig3} 
shows only the mass spectrum of the red-shifted lobe. Optically thin emission, LTE conditions, an excitation temperature of $T_{\rm ex}$=12~K \citep{Sanhueza2019,Li2020}, 
and a fractional abundance of CO relative to H$_{2}$ $f$(CO)=10$^{-4}$ were assumed in the calculations. 
The points exhibit significant spread but they indicate that the data can be described by a double power-law, 
$m$(v)$\propto$v$^{-\gamma}$, with $\gamma$=1.15$\pm$0.03 (Pearson's correlation coefficient $r$=$-$0.78) for v$_{\rm offset}$$<$50~km~s$^{-1}$ 
(LVC), and $\gamma$=1.75$\pm$0.20 ($r$=$-$0.40) for v$_{\rm offset}$$>$50~km~s$^{-1}$ (HVC). This double power-law is predicted by the 
numerical simulations of jet-driven bipolar outflows \citep{Smith1997}. The slopes of the power law of the outflow of core ALMA1 are slightly shallower than 
those of other sources \citep[e.g.,][]{Rodriguez1982}, but it has been found that the slope steepens with age and energy 
in the flow \citep{Richer2000}. Thus, the shallower slopes in Fig.~\ref{Fig3} would agree with the values expected for a young outflow embedded in a 70 $\mu$m dark 
massive clump at the early stages of high-mass star formation. Moreover, \cite{Solf1987} proposed a similar model to explain the spatio-kinematical properties 
of the jet associated with the source HH24-C in the low-mass HH24 complex. As \cite{Solf1987} mentions, the shape of the P-V diagram of HH24-C, obtained 
from one half of the outflow, resembles that of the Greek capital letter $\Lambda$. This is exactly the shape of the P-V diagram for each of the lobes   
of core ALMA1 (see Fig.~\ref{Fig2}b). In addition, from Fig.~\ref{Fig2}b it can be seen that the velocities at the tips of the lobes are the same for both HVC 
and LVC, strongly suggesting that there is a common mechanism  that drives them simultaneously. Given all these arguments, we conclude 
that the bipolar lobes seen in core ALMA1 are very likely due to the presence of a young jet-driven molecular outflow. 

Finally, we point out that, under this interpretation, the green points in Fig.~\ref{Fig2}b would correspond to the entrained material whose velocity field follows a 
Hubble-law, as it is seen in several other molecular outflows. Thus, the velocity offset at the origin of the outflow is expected to be close to zero, which justifies the 
definition of the velocity reference given in the previous section \S\ref{results}.

\subsection{Timescale of the molecular outflow}\label{age}

Fig.~\ref{Fig2}b shows the current configuration of positions and velocities for the different parts of the molecular outflow of core ALMA1.  
As \cite{Tafoya2019} pointed out, this type of plot only provides an instantaneous picture of the velocity field in the outflow at present time and 
cannot be used to trace the velocity history of the gas, unless the deceleration law as a function of time is known for each part of the outflow.
Nonetheless, despite of this limitation, important information of the kinematics of the outflow can be extracted by making some 
reasonable assumptions, which we describe in the following. From Fig.~\ref{Fig2}b it can be seen that the blue- and red-shifted parts of the HVC 
have material moving with a maximum velocity, v$_{\rm max, [B,R]}$ at the minimum position offset, r$_{\rm min, [B,R]}$, where the subscripts 
$\rm[B,R]$ stand for blue- and red-shifted part, respectively. Correspondingly, the minimum velocity, v$_{\rm min, [B,R]}$, is 
reached at the maximum position offset, r$_{\rm max, [B,R]}$. This means that, if one assumes that the driving jet has been launched with a 
constant velocity throughout its life, the material located at the tip of the jet has suffered a deceleration. Thus, using the size of the 
HVC and assuming a constant deceleration, the age of the outflow can be estimated. However, 
before attempting to estimate the age of the outflow, some considerations on its geometrical characteristics are necessary. 
 
 \begin{deluxetable*}{ccccccc}[!t]
 	\tablenum{1}
 	\tablecaption{Physical parameters for the kinematical components in the molecular outflow of core ALMA1\label{tab1}}
 	\tablewidth{0pt}
 	\tablehead{
 		\colhead{Component} & \colhead{$M$} & \colhead{$p$} & \colhead{$E_{\rm k}$} &\colhead{$\dot{M}$} & \colhead{$F_{\rm m}$} & \colhead{$L_{\rm m}$} \\
 		\colhead{Designation} & \colhead{$M_{\odot}$} & \colhead{$M_{\odot}$ km s$^{-1}$} & \colhead{ergs} &\colhead{$M_{\odot}$ yr$^{-1}$} & \colhead{$M_{\odot}$ km s$^{-1}$ yr$^{-1}$} & \colhead{$L_{\odot}$}
 	}
 	\startdata
 	HVC & 1.7-2.0$\times$10$^{-3}$& 1.1-1.3$\times$10$^{-1}$ & 7.1-8.5$\times$10$^{43}$& 1.9-2.3$\times$10$^{-6}$ & 1.2-1.5$\times$10$^{-4}$& 6.9-8.2$\times$10$^{-1}$ \\
 	LVC & 2.6-3.6$\times$10$^{-3}$ & 0.9-1.2$\times$10$^{-1}$& 3.2-4.2$\times$10$^{43}$& 3.0-4.2$\times$10$^{-6}$& 1.0-1.4$\times$10$^{-4}$ & 3.1-4.0$\times$10$^{-1}$ \\ 
 	\hline
 	Northern Lobe & 2.2-3.0$\times$10$^{-3}$& 1.0-1.2$\times$10$^{-1}$& 4.6-5.7$\times$10$^{43}$& 2.5-3.4$\times$10$^{-6}$& 1.1-1.4$\times$10$^{-4}$ & 4.3-5.3$\times$10$^{-1}$ \\
 	Southern Lobe & 2.0-2.7$\times$10$^{-3}$ & 1.0-1.2$\times$10$^{-1}$& 5.8-7.0$\times$10$^{43}$& 2.4-3.2$\times$10$^{-6}$& 1.2-1.5$\times$10$^{-4}$& 5.7-6.9$\times$10$^{-1}$ \\
 	\hline
 	Combined & 4.2-5.7$\times$10$^{-3}$& 2.0-2.4$\times$10$^{-1}$& 1.0-1.3$\times$10$^{44}$& 4.9-6.6$\times$10$^{-6}$ & 2.3-2.9$\times$10$^{-4}$ & 1.0-1.2 \\
 	\enddata
 	\tablecomments{The values were calculated assuming an inclination of the northern and southern lobes $\theta_{\rm R}$=54$^{\circ}$, $\theta_{\rm B}$=60$^{\circ}$, 
 		respectively, and an excitation temperature in the range $T_{\rm ex}$=10-30~K. The physical parameters are computed as 
 		follows: $M$=$\sum{M_{i}}$, $p$=$\sum{M_{i}{\rm v}_{i}}$, $E_{\rm k}$=$\sum{(M_{i}{\rm v}_{i}^{2}})/2$, $\dot{M}$=$\sum{M_{i}}$/$\Delta\tau$, $F_{\rm m}$=$p$/$\Delta\tau$ 
 		and $L_{\rm m}$=$E_{\rm k}$/$\Delta\tau$, where the subscript $i$ is the channel number and $\Delta\tau$ is the time that it takes to the gas to move from r$_{\rm min, [B,R]}$ to r$_{\rm max, [B,R]}$.}
 \end{deluxetable*}
 
One characteristic of the outflow of core ALMA1 that is readily seen from Fig.~\ref{Fig2} is that the northern and southern lobes have different 
spatial extents as well as different velocity offsets. While the maximum position offset for the southern lobe is 3$\rlap{.}^{\prime\prime}$7, it is 
only 2$\rlap{.}^{\prime\prime}$9 for the northern lobe. Similarly, the maximum 
velocity offsets for the southern and northern lobes are 78~km~s$^{-1}$ and 54~km~s$^{-1}$, respectively. 
However, these values are only projections of the position offset onto the plane of the sky and the velocity offset onto the line-of-sight 
direction, respectively, implying that the difference of size and velocity offset between the northern and southern lobes could be due to projection effects, 
i.e., one lobe may be more inclined than the other. Consequently, the inclination angles of the lobes, $[\theta_{\rm B}, \theta_{\rm R}]$, with respect to the plane-of-the-sky are needed to calculate the intrinsic values of the size and velocity offset (see Fig.~\ref{Fig4}a). In Appendix \S\ref{appendix} we show that it is possible to constrain 
the relative inclination of the lobes ($\theta_{\rm B}-\theta_{\rm R}$), which can then be used to further constrain their absolute inclination, by assuming that the 
launch velocity of the driving jet is the same for both lobes, i.e., v$_{\rm i, B}$=v$_{\rm i, R}$=v$_{\rm i}$, and that the blue- and redshifted parts of the jet have the 
same age, i.e., $\tau_{\rm jet, B}$=$\tau_{\rm jet, R}$=$\tau$. In the following subsection, \S\ref{energetics}, we provide arguments that support these assumptions.

Using the observed values of the projected sizes and line-of-sight velocities of the lobes, it is found that their relative inclination angle is restricted 
to the range 0$^{\circ}$$<$($\theta_{\rm B}-\theta_{\rm R}$)$<$9.5$^{\circ}$. Furthermore, for each given value of ($\theta_{\rm B}-\theta_{\rm R}$),  
there are only two pairs of inclination angles for the lobes. For example, if ($\theta_{\rm B}-\theta_{\rm R}$)=1$^{\circ}$, 
the inclination angles of the northern and southern lobes are either [$\theta_{\rm B}$=85$^{\circ}$, $\theta_{\rm R}$=84$^{\circ}$] or 
[$\theta_{\rm B}$=3$^{\circ}$, $\theta_{\rm R}$=2$^{\circ}$] (see Table~\ref{tab2}). This means that if we assume that the lobes are 
almost aligned with each other, their axis would lie basically either on the line-of-sight direction or on the plane-of-the-sky. However, this is 
an unlikely configuration since it would imply either unrealistic initial velocities and extremely short kinematical time scales, or too large sizes for the 
lobes (see Table~\ref{tab2}). On the other hand, if one considers an average inclination of the lobes equal to the mean inclination angle, 
($\theta_{\rm B}+\theta_{\rm R}$)/2=57.3$^{\circ}$ 
\citep[e.g.,][]{Bontemps1996,Beuther2002,Villiers2014}, equation~\ref{Eq:2} gives a relative angle of ($\theta_{\rm B}-\theta_{\rm R}$)$\approx$6$^{\circ}$ 
and the resulting age and size of the outflow are $\approx$1300~years and $\approx$0.2~pc, respectively (see Fig.~\ref{Fig4}b). 

\begin{figure*}
	\centering
	\includegraphics[angle=0,scale=0.7]{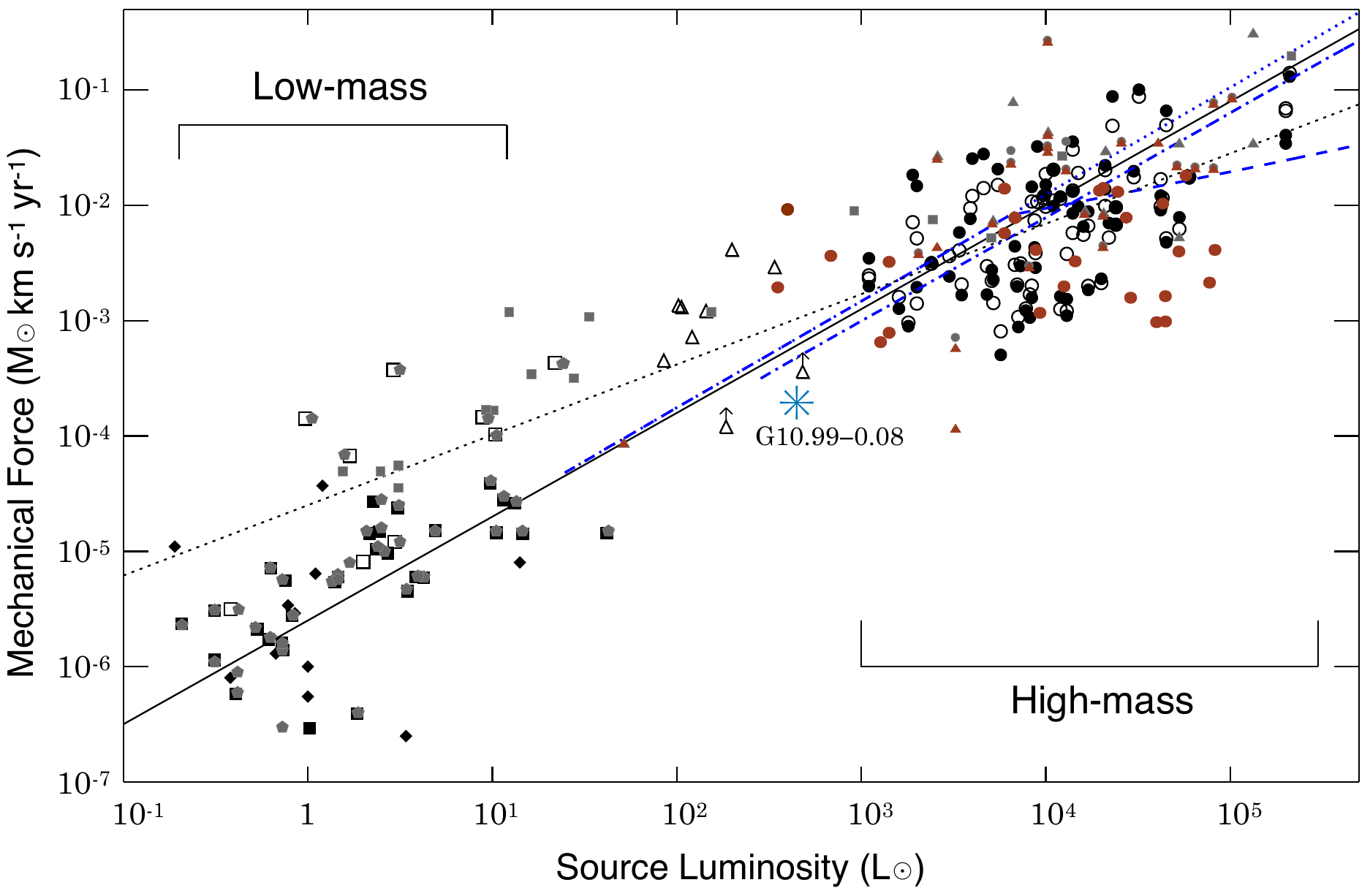}
	\caption{Mechanical force as a function of bolometric luminosity. The plot has been adapted from Figure 7 of \cite{Maud2015}, Figure 4 of \cite{Beuther2002} and  Figure 4 
	of \cite{Zhang2005}. The blue asterisk indicates the value for the outflow of core ALMA1.}
	\label{Fig4a}%
\end{figure*}

\subsection{Energetics of the molecular outflow}\label{energetics}

From Table~\ref{tab2} it can be seen that regardless of the relative inclination between the lobes, the final velocity of the northern lobe, 
v$_{\rm f, B}$, is lower than that of the southern lobe, v$_{\rm f, R}$. This difference could simply be indicating that the southern part of the jet 
is intrinsically faster than the northern part, or it could be indicating that the later is preferentially decelerated by ambient material. 
One way of investigating whether the initial velocities and ages for both parts of the jet are indeed equal or not is by means of comparing the momentum 
and mechanical force carried by the lobes. If the launch velocities and ages are the same for both parts of the jet, the momentum and mechanical 
force for both lobes should be the same too. 

In order to calculate the energetics of the molecular outflow of core ALMA1, we followed the same procedure as \cite{Tafoya2019}. Firstly, we
 assumed that all the velocity vectors within each lobe are parallel to the outflow axis, thus their de-projected magnitude is 
 v$=$v$_{\rm offset}/{\rm sin}\,\theta_{\rm inc}$. We adopted an inclination of the lobes $\theta_{\rm R}$=54$^{\circ}$, $\theta_{\rm B}$=60$^{\circ}$, 
 which, as discussed in the previous subsection, \S\ref{age}, results in an age of the outflow of $\tau_{\rm jet}$=1300~years. Subsequently, we calculated the 
 values of the physical parameters shown in Table~\ref{tab1} for each individual velocity channel, $i$, assuming LTE conditions, optically thin emission 
 and an excitation temperature in the range $T_{\rm ex}$=10-30~K. Finally, we added the values of the channels that correspond to a particular component, 
 namely northern lobe, southern lobe, HVC and LVC. The entrainment rate, $\dot{M}$, mechanical force, $F_{\rm m}$, and mechanical luminosity, 
 $L_{\rm m}$, were calculated using the times that it takes to the gas to move from r$_{\rm min, [B,R]}$ to r$_{\rm max, [B,R]}$, which are 
estimated to be 850 and 800 yr for the red- and blue-shifted lobe, respectively. The resulting values of the physical parameters are listed in Table~\ref{tab1}. 
The derived molecular mass for the outflow is lower than the one estimated by \cite{Pillai2019}, but their observations had lower angular resolution and were more 
affected by contamination of emission of the ambient material. As a matter of fact, in our analysis we do not consider the emission at the base of the outflow because 
of contamination of emission of the ambient material (see Fig.~\ref{Fig1b}b). Thus, the values of the molecular mass shown in Table~\ref{tab1} should be 
considered as lower limits. Nevertheless, using the mass \citep[$M$$\sim$10~$M_{\odot}$;][]{Sanhueza2019,Pillai2019} and bolometric luminosity 
($L_{\rm bol}$$\sim$470~$L_{\odot}$)\footnote{The luminosity is derived following equation 3 of \cite{Contreras2017}} for core ALMA1, and taking our estimation 
of the total mechanical force from Table~\ref{tab1}, which is a dynamical quantity that is not significantly affected by a poor sampling 
of the outflow, we find that core ALMA1 follows the mechanical force versus core mass and bolometric luminosity correlations obtained from molecular 
outflows in low- and high-mass star-forming regions (see Fig.\ref{Fig4a}). 
 
 In addition, the values of the momentum and mechanical force for the northern and southern lobes listed in Table~\ref{tab1} are, within the uncertainty 
 	range, basically the same. Thus, we conclude that, even though the final velocities of the northern and southern lobes are different, they must have been 
 ejected simultaneously and with the same initial velocity, as suggested in \S\ref{age}. 

\subsection{Nature of the deceleration of the HVC}\label{nature_jet}

There is now a plethora of observations of molecular outflows in low and high-mass star-forming regions that have revealed a rich variety of spatio-kinematical 
properties \citep[e.g.,][]{Bachiller1996, Arce2007, Kong2019, Zapata2019,Nony2020, Lee2020, Li2020}. As mentioned before, it is common that collimated molecular 
outflows exhibit components with a Hubble-law velocity profile and/or complex Hubble wedges, some of which may contain decelerating knots 
\cite[e.g.,][]{Bachiller1990,Plunkett2015,Nony2020}. Nevertheless, to the best of our knowledge, there has not been reported in the literature any other molecular 
outflow in a star-forming region whose P-V diagram has the $\mathsf{S}$-shaped morphology seen in G10.99$-$0.08. Particularly, the part of the P-V 
diagram that corresponds to the decelerating HVC, which we interpret as entrained material along the sides of the jet, is not seen in other outflows. One may wonder 
what is the origin of the deceleration of the HVC and what are the particular physical conditions of the molecular outflow of core ALMA1 that result in such a 
peculiar P-V diagram. The driving jet of the outflow, in principle, could be launched magnetocentrifugally, as it is proposed in the unified wind-driven model 
\citep{Shang2006}, although without the wide opening angle component (or with one that is too weak to be detected). Given the short age of the outflow of ALMA1 
($\sim$1000~years), it could be argued that the presence of the decelerating HVC occurs mainly at an early phase of the outflow when there is more material 
available to be entrained along the sides of the jet, i.e., the jet has not completely cleared out the inner parts of the lobes. However, similar time-scales have been 
estimated for outflows of cores with a wide range of masses \cite[1-100 $M_{\odot}$,][]{Nony2020} and they do not show the $\mathsf{S}$-shaped morphology 
seen in the outflow of core ALMA1. Consequently, the short age of the outflow does not seem to be the only factor for having such a P-V diagram. \cite{Chernin1994} 
found that a jet with Mach number $\leq$6 slows down rapidly by delivering energy and momentum as it entrains ambient material along its sides, which would 
account for the deceleration of the HVC. We thus propose that the observed deceleration of the HVC is likely due to the presence of a jet with a low Mach number that 
decelerates as it interacts with the ambient material. 
We also note that, given that the energetics of the outflow of core ALMA1 do not deviate from the trend seen in other outflows \citep{Bontemps1996,Beuther2002,Zhang2005}, 
future observations with high angular resolution should reveal more outflows with characteristics similar to that of G10.99$-$0.08.

On the other hand, we have so far assumed that the driving jet has been launched with a constant velocity throughout its life. However, an alternative possibility
that could explain the observed deceleration of the HVC is that the parts of the jet launched at earlier times (the ones further away from the protostar) were ejected at a lower 
velocity than those launched at later times (the ones closer to the protostar). This interpretation would imply that the jet is not ejected at a constant velocity, 
possibly shedding some light on the mass-accretion rate of the protostar, as we describe in the following. Since we do not know exactly the evolutionary stage of the 
protostar on the pre-main sequence track, we assume the accreting protostar is at ZAMS for the sake of estimating its mass. Considering the bolometric luminosity of the 
clump G10.99$-$0.08, $L_{\rm bol}$$\sim$470~$L_{\odot}$,  and ignoring the contribution of the accretion luminosity $L_{\rm acc}$, i.e., $L_{\rm acc}$ $<$ $L_{\star}$, 
which is valid for more massive stars with typical accretion rates, the ZAMS mass of the star would be $M_{\rm ZAMS}$$\approx$5~$M_{\odot}$ \citep{Schaller1992}. This 
implies that the current mass of the protostar is $\lesssim$5~$M_{\odot}$, assuming that all the luminosity from the clump is produced by a single protostar. Typically, 
one can consider that the initial velocity of the jet is comparable to the Keplerian velocity of the accretion disk at the jet launching location \cite{Pelletier1992}. 
Therefore, the increase in jet speed over time would imply an increase in dynamical mass of the central object over time, if assuming that 
the jet is launched at the same location \cite[e.g., see Figure 6 from][]{Rosen2020}. Based on the velocity differences seen in Fig.~\ref{Fig2}, the increase of 
dynamical mass, which is proportional to (v$_{\rm max}$/v$_{\rm min}$)$^{2}$, would be a factor of 2-3 times the initial mass. This would mean that there has been a significant 
increase in protostellar mass over the time scale of $\sim$1000~years, implying an upper limit for the mass-accretion rate of $\lesssim$2$\times$10$^{-3}$$M_{\odot}$~yr$^{-1}$. 
This value is rather large considering the physical parameters of the outflow listed in Table~\ref{tab1}, which makes it unlikely that the increase of the jet speed over 
time is due only to the increase in dynamical mass of the central object. Nonetheless, the combined effect of not having a constant launch velocity together and deceleration of 
the jet by the entrained material may explain the $\mathsf{S}$-shaped P-V diagram of this source.

In order to further explore these possibilities and to better understand the physical characteristics of molecular outflows that exhibit a decelerating HVC, observations with higher 
angular resolution, together with new models and numerical simulations, will be crucial. Recently, \cite{Tafoya2020} obtained high resolution images 
with ALMA of a jet-driven molecular outflow in the the evolved star W43A, which also exhibits an $\mathsf{S}$-shaped P-V diagram. The sharp images allowed them to clearly 
disentangle the different components of the molecular outflow, revealing an extremely collimated decelerating HVC. Future observations with higher angular resolution may 
allow us to unveil the structure of the molecular outflow of core ALMA1 too.

\section{Conclusions}\label{conclusions}

In this work we have carried out a spatio-kinematical analysis of the CO($J$=2$\rightarrow$1) line emission of the molecular outflow associated with 
the relatively massive core ALMA1 embedded in G10.99$-$0.08. The P-V diagram has a peculiar $\mathsf{S}$-shaped morphology that is explained by means 
of two components with different kinematical characteristics: an inner axial high-velocity component that moves with high velocity but decelerates, and a co-axial 
component moving with lower velocity but whose velocity increases with distance from the central source. The spatio-kinematical model is
interpreted as a jet-driven molecular outflow. The high-velocity decelerating component is associated with material moving near in the axis of the flow 
entrained by the underlying jet, or it could be the molecular component of the jet itself. The lower velocity component is interpreted as material lying 
in zones further away from the axis of the flow ambient but because of the entrainment by the inner gas it appears as if it was being accelerated. 
This interpretation is supported by the mass spectrum derived from the emission of the molecular gas. From an analysis of the P-V diagram profile we 
conclude that there is a relative angle between the axis of the the blue- and red-shifted lobes. Assuming an overall mean inclination angle of the 
outflow of $\approx$57$^{\circ}$, the relative angle between the lobes is $\approx$6$^{\circ}$. For such an inclination, the resulting age and size of 
the outflow are $\approx$1300~years and $\approx$0.2~pc, respectively. Further calculations and numerical simulations are necessary to better 
understand the physical characteristics of molecular outflows that have $\mathsf{S}$-shaped P-V diagrams. In addition, higher angular resolution 
observations are necessary to better constrain the input parameters of the models. 

Finally, we stress the fact that molecular outflows with very similar spatio-kinematical characteristics to those of the outflow of core ALMA1 are seen in 
evolved stars too. This shows that, despite being found in different astrophysical contexts, molecular outflows may be studied with a unified approach, 
which encourages us to develop more synergies between different fields of astronomy to broaden our knowledge on the physical processes 
underlying their formation and evolution.

\acknowledgments
P.S. was partially supported by a Grant-in-Aid for Scientific Research (KAKENHI Number 18H01259) of Japan Society for the Promotion of Science (JSPS). Y.C., 
and A.G. gratefully acknowledge the support from the NAOJ Visiting Fellow Program to visit the National Astronomical Observatory of Japan in 
November-December 2016. Data analysis was in part carried out on the Multi-wavelength Data Analysis System operated by the Astronomy Data Center (ADC), 
National Astronomical Observatory of Japan.  This paper makes use of the following ALMA dataset ADS/JAO.ALMA \#2015.1.01539.S. ALMA is a partnership of 
ESO (representing its member states), NSF (USA) and NINS (Japan), together with NRC (Canada) and NSC and ASIAA (Taiwan) and KASI (Republic of Korea), in 
co-operation with the Republic of Chile. The Joint ALMA Observatory is operated by ESO,AUI/NRAO and NAOJ. E de la F wishes to thank CUCEI and CUCEA, for 
financial support to visit Onsala Space Observatory, Chalmers University of Technology for several academic stays.

\appendix

\section{Appendix}\label{appendix}
  
\begin{figure*}
	\centering
	\includegraphics[angle=0,scale=0.8]{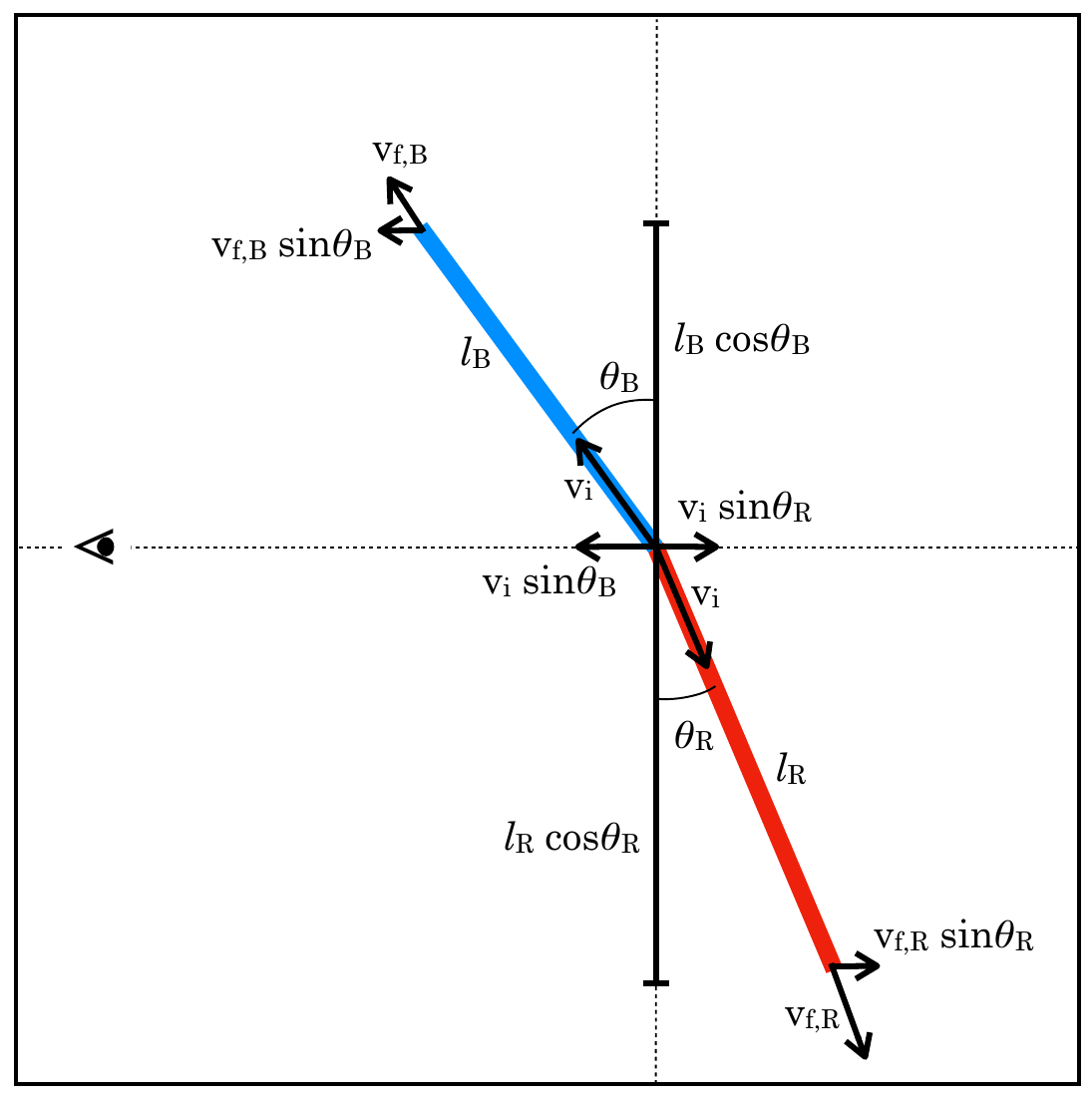}
	\caption{Diagram of the configuration of the blue- and red-shifted parts of the jet in the molecular outflow of G10.99$-$0.08. The initial 
		velocity of the jet, v$_{\rm i}$, is assumed to be the same for both parts. The observer is located on the left part of the diagram.}
	\label{Fig5}%
\end{figure*}

Consider the jet of core ALMA1, traced by the emission of the HVC, moving along a direction that has an inclination angle $\theta_{\rm R,B}$ with respect to the 
 plane of the sky, where the subscripts B and R indicate the blue- and red-shifted part of the jet, respectively (see Fig.~\ref{Fig5}). If both 
 parts of the jet were launched with the same initial velocity, v$_{\rm i}$, and the material at the tip of the jet has slowed down with a constant 
 deceleration to a final velocity, v$_{\rm f,[B,R]}$, over a distance, $l_{\rm B,R}$, then the travel-time (i.e., the age for each part of the jet), $\tau_{\rm jet [B,R]}$,  
 is given by 

  \begin{equation}\label{Eq:1}
  \tau_{\rm jet, [B,R]}=\frac{2\, l_{\rm B,R}}{{\rm v}_{\rm i}+{\rm v}_{\rm f,[B,R]}}.
  \end{equation}

Assuming that the blue- and red-shifted parts of the jet were launched simultaneously, i.e., $\tau_{\rm B} = \tau_{\rm R}$, and considering from 
Fig.~\ref{Fig2} that $l_{\rm B,R}={\rm r}_{\rm max,[B,R]}/{\rm cos}\,\theta_{\rm R,B}$ and v$_{\rm f,[B,R]}=$v$_{\rm min,[B,R]}/{\rm sin}\,\theta_{\rm R,B}$, 
one obtains the following equation:

 \begin{equation}\label{Eq:2}
 \frac{2\, {\rm r}_{\rm max, B}/{\rm cos}\,\theta_{\rm B}}{{\rm v}_{\rm i}+{\rm v}_{\rm min,B}/{\rm tan}\,\theta_{\rm B}} =  
 \frac{2\, {\rm r}_{\rm max, R}/{\rm cos}\,\theta_{\rm R}}{{\rm v}_{\rm i}+{\rm v}_{\rm min,R}/{\rm tan}\,\theta_{\rm R}}. 
 \end{equation}

Since the red-shifted part of the jet seems to have suffered less deceleration, the launch velocity can be approximated as 
v$_{\rm i}$$\approx$v$_{\rm max, R}/{\rm sin}\,\theta_{\rm R}$ and solve 
numerically $\theta_{\rm B}$=$\theta_{\rm B}(\theta_{\rm R})$ from equation~\ref{Eq:2}. Using the numerical values of  r$_{\rm max,[B,R]}$, 
v$_{\rm min,[B,R]}$ and v$_{\rm max, R}$ from Fig.~\ref{Fig2}, it is found that equation~\ref{Eq:2} has real solutions only for
0$^{\circ}<(\theta_{\rm B}-\theta_{\rm R})<9.5^{\circ}$. Therefore, it can be concluded that the relative inclination angle of the northern and 
southern lobes is less than~$\approx$9.5$^{\circ}$. Furthermore, for a given difference, ($\theta_{\rm B}-\theta_{\rm R}$), there are only two 
possible pairs of angles, $\theta_{\rm R,B}$, that solve Equation~\ref{Eq:2}. The values of $\theta_{\rm R,B}$ for given values of ($\theta_{\rm B}-\theta_{\rm R}$)
are listed in Table~\ref{tab2}. From Table~\ref{tab2} it can be seen that for $(\theta_{\rm B}-\theta_{\rm R})\approx0$, both $\theta_{\rm R,B}$ are 
either close to 0$^{\circ}$, which would imply an initial velocity for the jet of $\sim$2000~km~s$^{-1}$, or 90$^{\circ}$, which results in total length 
of the outflow of $\sim$0.6~pc. It is worth noting that for the relative angle $(\theta_{\rm B}-\theta_{\rm R})\approx9.5$$^{\circ}$ there is only one 
pair of angles that solves equation~\ref{Eq:2}, [$\theta_{\rm R}$=20$^{\circ}$ and $\theta_{\rm B}$=29.5$^{\circ}$], which  would imply an age of the 
outflow of $\tau=350$~years.

\begin{deluxetable*}{cccccccccc}[!h]
	\tablenum{2}
	\tablecaption{Inclination angle and spatio-kinematical parameters of the blue- red-shifted jets in core ALMA1\label{tab2}}
	\tablewidth{0pt}
	\tablehead{
		($\theta_{\rm B}-\theta_{\rm R}$) & Solution$^{1}$&$\theta_{\rm R}$ & $\theta_{\rm B}$ & v$_{\rm i}$ &v$_{\rm f,R}$ & v$_{\rm f,B}$&$l_{\rm R}$&$l_{\rm B}$&$\tau$\\
		degrees & &degrees & degrees & km s$^{-1}$ &km s$^{-1}$ & km s$^{-1}$& 10$^{17}$~cm & 10$^{17}$~cm & years}
	\startdata
	1 &Sol. 1 &84 & 85 & 80 & 48& 41& 17.7& 18.8& 9240\\
	 & Sol. 2&2 & 3 & 2292 & 1375& 783& 1.54& 1.96& 34 \\
	 \hline
	2 & Sol. 1&78 & 80 & 82 & 49& 42& 8.86& 9.43& 4569 \\
	 & Sol. 2&2 & 4 & 2292 & 1375& 588& 1.54& 1.96& 34 \\
	 \hline
	3 &Sol. 1& 72 & 75 & 84 & 50& 42& 5.95& 6.34& 2989 \\
	 &Sol. 2& 3 & 6 & 1529 & 917& 392& 1.55& 1.96& 51 \\
	 \hline
	4 &Sol. 1& 66 & 70 & 88 & 53& 44& 4.50& 4.82& 2181 \\
	 &Sol. 2& 4 & 8 & 1147 & 688& 295& 1.55& 1.96& 68 \\
	 \hline
	5 &Sol. 1 &60 & 65 & 92 & 55& 45& 3.64& 3.92& 1682 \\
	 &Sol. 2& 6 & 11 & 765 & 459& 215& 1.57& 1.97& 102 \\
	 \hline
	6 & Sol. 1&54 & 60 & 99 & 59& 47& 3.08& 3.33& 1337 \\
	 &Sol. 2& 7 & 13 & 656 & 394& 182& 1.58& 1.97& 119 \\
	 \hline 
	7 &Sol. 1& 48 & 55 & 108 & 65& 50& 2.68& 2.93& 1079 \\
	 & Sol. 2&9 & 16 & 511 & 307& 149& 1.60& 1.98& 154 \\
	 \hline
	8 & Sol. 1&41 & 49 & 122 & 73& 54& 2.35& 2.60& 844 \\
	 &Sol. 2& 11 & 19 & 419 & 252& 126& 1.63& 2.00& 189 \\
	 \hline
	9 &Sol. 1& 33 & 42 & 147 & 88& 61& 2.07& 2.34& 631 \\
	 &Sol. 2& 16 & 25 & 290 & 174& 97& 1.70& 2.04& 278 \\
	\enddata
	\tablecomments{$^{1}$For each value of ($\theta_{\rm B}-\theta_{\rm R}$) there are two possible combination of 
		solutions for $\theta_{\rm B}$ and $\theta_{\rm R}$ and the respective physical parameters derived from them.}
\end{deluxetable*}


\bibliography{bibliography_tafoya}{}

\begin{thebibliography}{}
\expandafter\ifx\csname natexlab\endcsname\relax\def\natexlab#1{#1}\fi

\bibitem[{{Arce} {et~al.}(2007){Arce}, {Shepherd}, {Gueth}, {Lee}, {Bachiller},
  {Rosen}, \& {Beuther}}]{Arce2007}
{Arce}, H.~G., {Shepherd}, D., {Gueth}, F., {et~al.} 2007, in Protostars and
  Planets V, ed. B.~{Reipurth}, D.~{Jewitt}, \& K.~{Keil} (Tuscon, AZ: Univ.
  Arizona Press), 245

\bibitem[{{Bachiller}(1996)}]{Bachiller1996}
{Bachiller}, R. 1996, \araa, 34, 111

\bibitem[{{Bachiller} {et~al.}(1990){Bachiller}, {Cernicharo},
  {Martin-Pintado}, {Tafalla}, \& {Lazareff}}]{Bachiller1990}
{Bachiller}, R., {Cernicharo}, J., {Martin-Pintado}, J., {Tafalla}, M., \&
  {Lazareff}, B. 1990, \aap, 231, 174

\bibitem[{{Bally}(2016)}]{Bally2016}
{Bally}, J. 2016, \araa, 54, 491

\bibitem[{{Banerjee} \& {Pudritz}(2006)}]{Banerjee2006}
{Banerjee}, R., \& {Pudritz}, R.~E. 2006, \apj, 641, 949

\bibitem[{{Beuther} {et~al.}(2002){Beuther}, {Schilke}, {Sridharan}, {Menten},
  {Walmsley}, \& {Wyrowski}}]{Beuther2002}
{Beuther}, H., {Schilke}, P., {Sridharan}, T.~K., {et~al.} 2002, \aap, 383, 892

\bibitem[{{Bontemps} {et~al.}(1996){Bontemps}, {Andre}, {Terebey}, \&
  {Cabrit}}]{Bontemps1996}
{Bontemps}, S., {Andre}, P., {Terebey}, S., \& {Cabrit}, S. 1996, \aap, 311,
  858

\bibitem[{{Cabrit} \& {Bertout}(1986)}]{Cabrit1986}
{Cabrit}, S., \& {Bertout}, C. 1986, \apj, 307, 313

\bibitem[{{Cabrit} \& {Bertout}(1990)}]{Cabrit1990}
---. 1990, \apj, 348, 530

\bibitem[{{Cabrit} {et~al.}(1997){Cabrit}, {Raga}, \& {Gueth}}]{Cabrit1997}
{Cabrit}, S., {Raga}, A., \& {Gueth}, F. 1997, in IAU Symposium, Vol. 182,
  Herbig-Haro Flows and the Birth of Stars, ed. B.~{Reipurth} \& C.~{Bertout},
  163--180

\bibitem[{{Carey} {et~al.}(2009){Carey}, {Noriega-Crespo}, {Mizuno}, {Shenoy},
  {Paladini}, {Kraemer}, {Price}, {Flagey}, {Ryan}, {Ingalls}, {Kuchar},
  {Pinheiro Gon{\c{c}}alves}, {Indebetouw}, {Billot}, {Marleau}, {Padgett},
  {Rebull}, {Bressert}, {Ali}, {Molinari}, {Martin}, {Berriman}, {Boulanger},
  {Latter}, {Miville-Deschenes}, {Shipman}, \& {Testi}}]{Carey2009}
{Carey}, S.~J., {Noriega-Crespo}, A., {Mizuno}, D.~R., {et~al.} 2009, \pasp,
  121, 76

\bibitem[{{Chernin} {et~al.}(1994){Chernin}, {Masson}, {Gouveia dal Pino}, \&
  {Benz}}]{Chernin1994}
{Chernin}, L., {Masson}, C., {Gouveia dal Pino}, E.~M., \& {Benz}, W. 1994,
  \apj, 426, 204

\bibitem[{{Churchwell} {et~al.}(2009){Churchwell}, {Babler}, {Meade},
  {Whitney}, {Benjamin}, {Indebetouw}, {Cyganowski}, {Robitaille}, {Povich},
  {Watson}, \& {Bracker}}]{Churchwell2009}
{Churchwell}, E., {Babler}, B.~L., {Meade}, M.~R., {et~al.} 2009, \pasp, 121,
  213

\bibitem[{{Contreras} {et~al.}(2017){Contreras}, {Rathborne}, {Guzman},
  {Jackson}, {Whitaker}, {Sanhueza}, \& {Foster}}]{Contreras2017}
{Contreras}, Y., {Rathborne}, J.~M., {Guzman}, A., {et~al.} 2017, \mnras, 466,
  340

\bibitem[{{Contreras} {et~al.}(2013){Contreras}, {Schuller}, {Urquhart},
  {Csengeri}, {Wyrowski}, {Beuther}, {Bontemps}, {Bronfman}, {Henning},
  {Menten}, {Schilke}, {Walmsley}, {Wienen}, {Tackenberg}, \&
  {Linz}}]{Contreras2013}
{Contreras}, Y., {Schuller}, F., {Urquhart}, J.~S., {et~al.} 2013, \aap, 549,
  A45

\bibitem[{{Contreras} {et~al.}(2018){Contreras}, {Sanhueza}, {Jackson},
  {Guzm{\'a}n}, {Longmore}, {Garay}, {Zhang}, {Nguyễn-Lu'o'ng}, {Tatematsu},
  {Nakamura}, {Sakai}, {Ohashi}, {Liu}, {Saito}, {Gomez}, {Rathborne}, \&
  {Whitaker}}]{Contreras2018a}
{Contreras}, Y., {Sanhueza}, P., {Jackson}, J.~M., {et~al.} 2018, \apj, 861, 14

\bibitem[{{de Villiers} {et~al.}(2014){de Villiers}, {Chrysostomou},
  {Thompson}, {Ellingsen}, {Urquhart}, {Breen}, {Burton}, {Csengeri}, \&
  {Ward-Thompson}}]{Villiers2014}
{de Villiers}, H.~M., {Chrysostomou}, A., {Thompson}, M.~A., {et~al.} 2014,
  \mnras, 444, 566

\bibitem[{{De Young}(1986)}]{De-Young1986}
{De Young}, D.~S. 1986, \apj, 307, 62

\bibitem[{{Foster} {et~al.}(2011){Foster}, {Jackson}, {Barnes}, {Barris},
  {Brooks}, {Cunningham}, {Finn}, {Fuller}, {Longmore}, {Mascoop}, {Peretto},
  {Rathborne}, {Sanhueza}, {Schuller}, \& {Wyrowski}}]{Foster2011}
{Foster}, J.~B., {Jackson}, J.~M., {Barnes}, P.~J., {et~al.} 2011, \apjs, 197,
  25

\bibitem[{{Foster} {et~al.}(2013){Foster}, {Rathborne}, {Sanhueza},
  {Claysmith}, {Whitaker}, {Jackson}, {Mascoop}, {Wienen}, {Breen}, {Herpin},
  {Duarte-Cabral}, {Csengeri}, {Contreras}, {Indermuehle}, {Barnes}, {Walsh},
  {Cunningham}, {Britton}, {Voronkov}, {Urquhart}, {Alves}, {Jordan}, {Hill},
  {Hoq}, {Brooks}, \& {Longmore}}]{Foster2013}
{Foster}, J.~B., {Rathborne}, J.~M., {Sanhueza}, P., {et~al.} 2013, \pasa, 30,
  e038

\bibitem[{{Guzm{\'a}n} {et~al.}(2015){Guzm{\'a}n}, {Sanhueza}, {Contreras},
  {Smith}, {Jackson}, {Hoq}, \& {Rathborne}}]{Guzman2015}
{Guzm{\'a}n}, A.~E., {Sanhueza}, P., {Contreras}, Y., {et~al.} 2015, \apj, 815,
  130

\bibitem[{{Henning} {et~al.}(2010){Henning}, {Linz}, {Krause}, {Ragan},
  {Beuther}, {Launhardt}, {Nielbock}, \& {Vasyunina}}]{Henning2010}
{Henning}, T., {Linz}, H., {Krause}, O., {et~al.} 2010, \aap, 518, L95

\bibitem[{{Hirano} {et~al.}(2010){Hirano}, {Ho}, {Liu}, {Shang}, {Lee}, \&
  {Bourke}}]{Hirano2010}
{Hirano}, N., {Ho}, P.~P.~T., {Liu}, S.-Y., {et~al.} 2010, \apj, 717, 58

\bibitem[{{Hirano} {et~al.}(2006){Hirano}, {Liu}, {Shang}, {Ho}, {Huang},
  {Kuan}, {McCaughrean}, \& {Zhang}}]{Hirano2006}
{Hirano}, N., {Liu}, S.-Y., {Shang}, H., {et~al.} 2006, \apjl, 636, L141

\bibitem[{{Jackson} {et~al.}(2013){Jackson}, {Rathborne}, {Foster}, {Whitaker},
  {Sanhueza}, {Claysmith}, {Mascoop}, {Wienen}, {Breen}, {Herpin},
  {Duarte-Cabral}, {Csengeri}, {Longmore}, {Contreras}, {Indermuehle},
  {Barnes}, {Walsh}, {Cunningham}, {Brooks}, {Britton}, {Voronkov}, {Urquhart},
  {Alves}, {Jordan}, {Hill}, {Hoq}, {Finn}, {Bains}, {Bontemps}, {Bronfman},
  {Caswell}, {Deharveng}, {Ellingsen}, {Fuller}, {Garay}, {Green}, {Hindson},
  {Jones}, {Lenfestey}, {Lo}, {Lowe}, {Mardones}, {Menten}, {Minier}, {Morgan},
  {Motte}, {Muller}, {Peretto}, {Purcell}, {Schilke}, {Bontemps}, {Schuller},
  {Titmarsh}, {Wyrowski}, \& {Zavagno}}]{Jackson2013}
{Jackson}, J.~M., {Rathborne}, J.~M., {Foster}, J.~B., {et~al.} 2013, \pasa,
  30, e057

\bibitem[{{Kainulainen} {et~al.}(2013){Kainulainen}, {Ragan}, {Henning}, \&
  {Stutz}}]{Kainulainen2013}
{Kainulainen}, J., {Ragan}, S.~E., {Henning}, T., \& {Stutz}, A. 2013, \aap,
  557, A120

\bibitem[{{Kong} {et~al.}(2019){Kong}, {Arce}, {Maureira}, {Caselli}, {Tan}, \&
  {Fontani}}]{Kong2019}
{Kong}, S., {Arce}, H.~G., {Maureira}, M.~J., {et~al.} 2019, \apj, 874, 104

\bibitem[{{Lada} \& {Fich}(1996)}]{Lada1996}
{Lada}, C.~J., \& {Fich}, M. 1996, \apj, 459, 638

\bibitem[{{Lee}(2020)}]{Lee2020}
{Lee}, C.-F. 2020, \aapr, 28, 1

\bibitem[{{Li} {et~al.}(2019{\natexlab{a}}){Li}, {Zhang}, {Pillai}, {Stephens},
  {Wang}, \& {Li}}]{Li2019}
{Li}, S., {Zhang}, Q., {Pillai}, T., {et~al.} 2019{\natexlab{a}}, \apj, 886,
  130

\bibitem[{{Li} {et~al.}(2019{\natexlab{b}}){Li}, {Wang}, {Fang}, {Zhang}, {Li},
  {Zhang}, {Li}, {Zhu}, \& {Zeng}}]{Li2019a}
{Li}, S., {Wang}, J., {Fang}, M., {et~al.} 2019{\natexlab{b}}, \apj, 878, 29

\bibitem[{{Li} {et~al.}(2020){Li}, {Sanhueza}, {Zhang}, {Nakamura}, {Lu},
  {Wang}, {Liu}, {Tatematsu}, {Jackson}, {Silva}, {Guzm{\'a}n}, {Sakai},
  {Izumi}, {Tafoya}, {Li}, {Contreras}, {Morii}, \& {Kim}}]{Li2020}
{Li}, S., {Sanhueza}, P., {Zhang}, Q., {et~al.} 2020, \apj, 903, 119

\bibitem[{{Lu} {et~al.}(2015){Lu}, {Zhang}, {Wang}, \& {Gu}}]{Lu2015}
{Lu}, X., {Zhang}, Q., {Wang}, K., \& {Gu}, Q. 2015, \apj, 805, 171

\bibitem[{{Machida} {et~al.}(2008){Machida}, {Inutsuka}, \&
  {Matsumoto}}]{Machida2008}
{Machida}, M.~N., {Inutsuka}, S.-i., \& {Matsumoto}, T. 2008, \apj, 676, 1088

\bibitem[{{Masson} \& {Chernin}(1992)}]{Masson1992}
{Masson}, C.~R., \& {Chernin}, L.~M. 1992, \apjl, 387, L47

\bibitem[{{Maud} {et~al.}(2015){Maud}, {Moore}, {Lumsden}, {Mottram},
  {Urquhart}, \& {Hoare}}]{Maud2015}
{Maud}, L.~T., {Moore}, T.~J.~T., {Lumsden}, S.~L., {et~al.} 2015, \mnras, 453,
  645

\bibitem[{{Molinari} {et~al.}(2010){Molinari}, {Swinyard}, {Bally}, {Barlow},
  {Bernard}, {Martin}, {Moore}, {Noriega-Crespo}, {Plume}, {Testi}, {Zavagno},
  {Abergel}, {Ali}, {Andr{\'e}}, {Baluteau}, {Benedettini}, {Bern{\'e}},
  {Billot}, {Blommaert}, {Bontemps}, {Boulanger}, {Brand}, {Brunt}, {Burton},
  {Campeggio}, {Carey}, {Caselli}, {Cesaroni}, {Cernicharo}, {Chakrabarti},
  {Chrysostomou}, {Codella}, {Cohen}, {Compiegne}, {Davis}, {de Bernardis}, {de
  Gasperis}, {Di Francesco}, {di Giorgio}, {Elia}, {Faustini}, {Fischera},
  {Fukui}, {Fuller}, {Ganga}, {Garcia-Lario}, {Giard}, {Giardino}, {Glenn},
  {Goldsmith}, {Griffin}, {Hoare}, {Huang}, {Jiang}, {Joblin}, {Joncas},
  {Juvela}, {Kirk}, {Lagache}, {Li}, {Lim}, {Lord}, {Lucas}, {Maiolo},
  {Marengo}, {Marshall}, {Masi}, {Massi}, {Matsuura}, {Meny}, {Minier},
  {Miville-Desch{\^e}nes}, {Montier}, {Motte}, {M{\"u}ller}, {Natoli}, {Neves},
  {Olmi}, {Paladini}, {Paradis}, {Pestalozzi}, {Pezzuto}, {Piacentini},
  {Pomar{\`e}s}, {Popescu}, {Reach}, {Richer}, {Ristorcelli}, {Roy}, {Royer},
  {Russeil}, {Saraceno}, {Sauvage}, {Schilke}, {Schneider-Bontemps},
  {Schuller}, {Schultz}, {Shepherd}, {Sibthorpe}, {Smith}, {Smith},
  {Spinoglio}, {Stamatellos}, {Strafella}, {Stringfellow}, {Sturm}, {Taylor},
  {Thompson}, {Tuffs}, {Umana}, {Valenziano}, {Vavrek}, {Viti}, {Waelkens},
  {Ward-Thompson}, {White}, {Wyrowski}, {Yorke}, \& {Zhang}}]{Molinari2010}
{Molinari}, S., {Swinyard}, B., {Bally}, J., {et~al.} 2010, \pasp, 122, 314

\bibitem[{{Nony} {et~al.}(2020){Nony}, {Motte}, {Louvet}, {Plunkett},
  {Gusdorf}, {Fechtenbaum}, {Pouteau}, {Lefloch}, {Bontemps}, {Molet}, \&
  {Robitaille}}]{Nony2020}
{Nony}, T., {Motte}, F., {Louvet}, F., {et~al.} 2020, \aap, 636, A38

\bibitem[{{Palau} {et~al.}(2006){Palau}, {Ho}, {Zhang}, {Estalella}, {Hirano},
  {Shang}, {Lee}, {Bourke}, {Beuther}, \& {Kuan}}]{Palau2006}
{Palau}, A., {Ho}, P.~T.~P., {Zhang}, Q., {et~al.} 2006, \apjl, 636, L137

\bibitem[{{Pelletier} \& {Pudritz}(1992)}]{Pelletier1992}
{Pelletier}, G., \& {Pudritz}, R.~E. 1992, \apj, 394, 117

\bibitem[{{Pillai} {et~al.}(2019){Pillai}, {Kauffmann}, {Zhang}, {Sanhueza},
  {Leurini}, {Wang}, {Sridharan}, \& {König}}]{Pillai2019}
{Pillai}, T., {Kauffmann}, J., {Zhang}, Q., {et~al.} 2019, \aap, 622, A54

\bibitem[{{Pillai} {et~al.}(2006){Pillai}, {Wyrowski}, {Menten}, \&
  {Kr{\"u}gel}}]{Pillai2006}
{Pillai}, T., {Wyrowski}, F., {Menten}, K.~M., \& {Kr{\"u}gel}, E. 2006, \aap,
  447, 929

\bibitem[{{Plunkett} {et~al.}(2015){Plunkett}, {Arce}, {Mardones}, {van
  Dokkum}, {Dunham}, {Fernández-López}, {Gallardo}, \&
  {Corder}}]{Plunkett2015}
{Plunkett}, A.~L., {Arce}, H.~G., {Mardones}, D., {et~al.} 2015, \nat, 527, 70

\bibitem[{{Qiu} {et~al.}(2008){Qiu}, {Zhang}, {Megeath}, {Gutermuth},
  {Beuther}, {Shepherd}, {Sridharan}, {Testi}, \& {De Pree}}]{Qiu2008}
{Qiu}, K., {Zhang}, Q., {Megeath}, S.~T., {et~al.} 2008, \apj, 685, 1005

\bibitem[{{Raga} \& {Cabrit}(1993)}]{Raga1993}
{Raga}, A., \& {Cabrit}, S. 1993, \aap, 278, 267

\bibitem[{{Rathborne} {et~al.}(2016){Rathborne}, {Whitaker}, {Jackson},
  {Foster}, {Contreras}, {Stephens}, {Guzm{\'a}n}, {Longmore}, {Sanhueza},
  {Schuller}, {Wyrowski}, \& {Urquhart}}]{Rathborne2016}
{Rathborne}, J.~M., {Whitaker}, J.~S., {Jackson}, J.~M., {et~al.} 2016, \pasa,
  33, e030

\bibitem[{{Richer} {et~al.}(2000){Richer}, {Shepherd}, {Cabrit}, {Bachiller},
  \& {Churchwell}}]{Richer2000}
{Richer}, J.~S., {Shepherd}, D.~S., {Cabrit}, S., {Bachiller}, R., \&
  {Churchwell}, E. 2000, in Protostars and Planets IV, ed. V.~{Mannings}, A.~P.
  {Boss}, \& S.~S. {Russell} (Tuscon, AZ: Univ. Arizona Press), 867

\bibitem[{{Rodriguez} {et~al.}(1982){Rodriguez}, {Carral}, {Ho}, \&
  {Moran}}]{Rodriguez1982}
{Rodriguez}, L.~F., {Carral}, P., {Ho}, P.~T.~P., \& {Moran}, J.~M. 1982, \apj,
  260, 635

\bibitem[{{Rosen} \& {Krumholz}(2020)}]{Rosen2020}
{Rosen}, A.~L., \& {Krumholz}, M.~R. 2020, \aj, 160, 78

\bibitem[{{Sahai} {et~al.}(2017){Sahai}, {Vlemmings}, {Gledhill}, {Sánchez
  Contreras}, {Lagadec}, {Nyman}, \& {Quintana-Lacaci}}]{Sahai2017}
{Sahai}, R., {Vlemmings}, W.~H.~T., {Gledhill}, T., {et~al.} 2017, \apjl, 835,
  L13

\bibitem[{{Sakai} {et~al.}(2013){Sakai}, {Sakai}, {Foster}, {Sanhueza},
  {Jackson}, {Kassis}, {Furuya}, {Aikawa}, {Hirota}, \& {Yamamoto}}]{Sakai2013}
{Sakai}, T., {Sakai}, N., {Foster}, J.~B., {et~al.} 2013, \apjl, 775, L31

\bibitem[{{Sanhueza} {et~al.}(2010){Sanhueza}, {Garay}, {Bronfman}, {Mardones},
  {May}, \& {Saito}}]{Sanhueza2010}
{Sanhueza}, P., {Garay}, G., {Bronfman}, L., {et~al.} 2010, \apj, 715, 18

\bibitem[{{Sanhueza} {et~al.}(2012){Sanhueza}, {Jackson}, {Foster}, {Garay},
  {Silva}, \& {Finn}}]{Sanhueza2012}
{Sanhueza}, P., {Jackson}, J.~M., {Foster}, J.~B., {et~al.} 2012, \apj, 756, 60

\bibitem[{{Sanhueza} {et~al.}(2013){Sanhueza}, {Jackson}, {Foster},
  {Jimenez-Serra}, {Dirienzo}, \& {Pillai}}]{Sanhueza2013}
---. 2013, \apj, 773, 123

\bibitem[{{Sanhueza} {et~al.}(2017){Sanhueza}, {Jackson}, {Zhang},
  {Guzm{\'a}n}, {Lu}, {Stephens}, {Wang}, \& {Tatematsu}}]{Sanhueza2017}
{Sanhueza}, P., {Jackson}, J.~M., {Zhang}, Q., {et~al.} 2017, \apj, 841, 97

\bibitem[{{Sanhueza} {et~al.}(2019){Sanhueza}, {Contreras}, {Wu}, {Jackson},
  {Guzm{\'a}n}, {Zhang}, {Li}, {Lu}, {Silva}, {Izumi}, {Liu}, {Miura},
  {Tatematsu}, {Sakai}, {Beuther}, {Garay}, {Ohashi}, {Saito}, {Nakamura},
  {Saigo}, {Veena}, {Nguyen-Luong}, \& {Tafoya}}]{Sanhueza2019}
{Sanhueza}, P., {Contreras}, Y., {Wu}, B., {et~al.} 2019, \apj, 886, 102

\bibitem[{{Schaller} {et~al.}(1992){Schaller}, {Schaerer}, {Meynet}, \&
  {Maeder}}]{Schaller1992}
{Schaller}, G., {Schaerer}, D., {Meynet}, G., \& {Maeder}, A. 1992, \aaps, 96,
  269

\bibitem[{{Schuller} {et~al.}(2009){Schuller}, {Menten}, {Contreras},
  {Wyrowski}, {Schilke}, {Bronfman}, {Henning}, {Walmsley}, {Beuther},
  {Bontemps}, {Cesaroni}, {Deharveng}, {Garay}, {Herpin}, {Lefloch}, {Linz},
  {Mardones}, {Minier}, {Molinari}, {Motte}, {Nyman}, {Reveret}, {Risacher},
  {Russeil}, {Schneider}, {Testi}, {Troost}, {Vasyunina}, {Wienen}, {Zavagno},
  {Kovacs}, {Kreysa}, {Siringo}, \& {Wei{\ss}}}]{Schuller2009}
{Schuller}, F., {Menten}, K.~M., {Contreras}, Y., {et~al.} 2009, \aap, 504, 415

\bibitem[{{Shang} {et~al.}(2006){Shang}, {Allen}, {Li}, {Liu}, {Chou}, \&
  {Anderson}}]{Shang2006}
{Shang}, H., {Allen}, A., {Li}, Z.-Y., {et~al.} 2006, \apj, 649, 845

\bibitem[{{Shu} {et~al.}(1991){Shu}, {Ruden}, {Lada}, \& {Lizano}}]{Shu1991}
{Shu}, F.~H., {Ruden}, S.~P., {Lada}, C.~J., \& {Lizano}, S. 1991, \apjl, 370,
  L31

\bibitem[{{Smith} {et~al.}(1997){Smith}, {Suttner}, \& {Yorke}}]{Smith1997}
{Smith}, M.~D., {Suttner}, G., \& {Yorke}, H.~W. 1997, \aap, 323, 223

\bibitem[{{Solf}(1987)}]{Solf1987}
{Solf}, J. 1987, \aap, 184, 322

\bibitem[{{Svoboda} {et~al.}(2019){Svoboda}, {Shirley}, {Traficante},
  {Battersby}, {Fuller}, {Zhang}, {Beuther}, {Peretto}, {Brogan}, \&
  {Hunter}}]{Svoboda2019}
{Svoboda}, B.~E., {Shirley}, Y.~L., {Traficante}, A., {et~al.} 2019, \apj, 886,
  36

\bibitem[{{Tafoya} {et~al.}(2020){Tafoya}, {Imai}, {G{\'o}mez}, {Nakashima},
  {Orosz}, \& {Yung}}]{Tafoya2020}
{Tafoya}, D., {Imai}, H., {G{\'o}mez}, J.~F., {et~al.} 2020, \apjl, 890, L14

\bibitem[{{Tafoya} {et~al.}(2019){Tafoya}, {Orosz}, {Vlemmings}, {Sahai}, \&
  {P\'erez-S\'anchez}}]{Tafoya2019}
{Tafoya}, D., {Orosz}, G., {Vlemmings}, W.~H.~T., {Sahai}, R., \&
  {P\'erez-S\'anchez}, A.~F. 2019, \aap, 629, A8

\bibitem[{{Tan} {et~al.}(2013){Tan}, {Kong}, {Butler}, {Caselli}, \&
  {Fontani}}]{Tan2013}
{Tan}, J.~C., {Kong}, S., {Butler}, M.~J., {Caselli}, P., \& {Fontani}, F.
  2013, \apj, 779, 96

\bibitem[{{Wang} {et~al.}(2016){Wang}, {Testi}, {Burkert}, {Walmsley},
  {Beuther}, \& {Henning}}]{Wang2016}
{Wang}, K., {Testi}, L., {Burkert}, A., {et~al.} 2016, \apjs, 226, 9

\bibitem[{{Wang} {et~al.}(2011){Wang}, {Zhang}, {Wu}, \& {Zhang}}]{Wang2011}
{Wang}, K., {Zhang}, Q., {Wu}, Y., \& {Zhang}, H. 2011, \apj, 735, 64

\bibitem[{{Whitaker} {et~al.}(2017){Whitaker}, {Jackson}, {Rathborne},
  {Foster}, {Contreras}, {Sanhueza}, {Stephens}, \& {Longmore}}]{Whitaker2017}
{Whitaker}, J.~S., {Jackson}, J.~M., {Rathborne}, J.~M., {et~al.} 2017, \aj,
  154, 140

\bibitem[{{Yang} {et~al.}(2018){Yang}, {Thompson}, {Urquhart}, \&
  {Tian}}]{Yang2018}
{Yang}, A.~Y., {Thompson}, M.~A., {Urquhart}, J.~S., \& {Tian}, W.~W. 2018,
  \apjs, 235, 3

\bibitem[{{Zapata} {et~al.}(2019){Zapata}, {Ho}, {Guzm{\'a}n Ccolque},
  {Fern{\'a}ndez-Lop{\'e}z}, {Rodr{\'\i}guez}, {Bally}, {Sanhueza}, {Palau}, \&
  {Saito}}]{Zapata2019}
{Zapata}, L.~A., {Ho}, P. T.~P., {Guzm{\'a}n Ccolque}, E., {et~al.} 2019,
  \mnras, 486, L15

\bibitem[{{Zhang} {et~al.}(2005){Zhang}, {Hunter}, {Brand}, {Sridharan},
  {Cesaroni}, {Molinari}, {Wang}, \& {Kramer}}]{Zhang2005}
{Zhang}, Q., {Hunter}, T.~R., {Brand}, J., {et~al.} 2005, \apj, 625, 864

\bibitem[{{Zhang} {et~al.}(2015){Zhang}, {Wang}, {Lu}, \&
  {Jim{\'e}nez-Serra}}]{Zhang2015}
{Zhang}, Q., {Wang}, K., {Lu}, X., \& {Jim{\'e}nez-Serra}, I. 2015, \apj, 804,
  141

\end{thebibliography}
\bibliographystyle{aasjournal}

\end{CJK*}
\end{document}